\documentclass[aps,prc,twocolumn,nofootinbib]{revtex4-1} %nofootinbib
\usepackage{epsfig}
\usepackage{color}
\usepackage[latin1]{inputenc}
\usepackage{float,amsmath}
\usepackage{graphicx}
\usepackage{tikz}
\usepackage[percent]{overpic}

\newcommand{\xt}{{\mathbf{x}_\perp}}
\newcommand{\kt}{{\mathbf{k}_\perp}}
\newcommand{\yt}{{\mathbf{y}_\perp}}

\begin{document}

\title{3-D Glasma initial state for relativistic heavy ion collisions}

\author{Bj\"orn Schenke}
\affiliation{Physics Department, Brookhaven National Laboratory, Upton, NY 11973, USA}
\author{S\"oren Schlichting}
\affiliation{Physics Department, Brookhaven National Laboratory, Upton, NY 11973, USA}

\begin{abstract}
We extend the impact parameter dependent Glasma model (IP-Glasma) to three dimensions using explicit small $x$ evolution of the 
two incoming nuclear gluon distributions. We compute rapidity distributions of produced gluons
and the early time energy momentum tensor as a function of space-time rapidity and transverse coordinates. 
We study rapidity correlations and fluctuations of the initial geometry and multiplicity distributions and
compare to existing models for the three dimensional initial state.
\end{abstract}

%\pacs{11.15Bt, 04.25.Nx, 11.10Wx, 12.38Mh}
\maketitle

%%%%%%%%%%%%%%%%%%%%%%%%%%%%%%%%%%%%%%%%%%%%%%%%%%%%%%%%%%%%%%%

\section{Introduction}
Sophisticated simulations of the space-time evolution of a heavy-ion collision using relativistic viscous hydrodynamics have served as a powerful tool to constrain the properties of the Quark Gluon Plasma (QGP), a new state of matter created in high-energy nucleus-nucleus collisions at the Relativistic Heavy Ion Collider (RHIC) and the Large Hadron Collider (LHC) \cite{Gale:2013da}. While an improved theoretical understanding of the initial state geometry has been essential to this progress \cite{Alver:2010gr,Schenke:2011bn,Gale:2012rq}, a significant uncertainty still surrounds our knowledge of the three dimensional event-by-event geometry of the initial state. While it has been popular to assume that the space-time evolution of high-energy collisions can be described as approximately boost invariant, even at energies available at the LHC it is desirable to loosen this constraint and explore the full three-dimensional dynamics of the system. Naturally, this is expected to yield novel insights into the properties of the QGP. For example, it was recently shown that  at RHIC energies the rapidity dependence of flow harmonics can reveal additional information on the medium properties such as the temperature dependence of the shear viscosity to entropy density ratio $\eta/s$ \cite{Denicol:2015nhu}. 

The results of hydrodynamic simulations are very sensitive to the initial condition for the energy momentum tensor. Yet, our current understanding of the initial conditions for heavy ion collisions in three spatial dimensions is very limited. So far calculations within effective field theories of high-energy QCD only provide boost invariant (effectively two dimensional) initial conditions \cite{Schenke:2012wb,Schenke:2012hg}, while models that do determine three dimensional initial conditions either greatly simplify the rapidity dependence \cite{Hirano:2001eu,Nonaka:2006yn,Schenke:2010rr,Bozek:2011ua,Molnar:2014zha}, or are based on hadronic degrees of freedom \cite{Steinheimer:2007iy} and thus should not directly apply at high-energies. Even though some more recent models include the notion of ``color flux tubes'' of variable length or other fluctuations in rapidity \cite{Werner:2010aa,Monnai:2015sca,Bozek:2015tca,Denicol:2015nhu}, the rapidity dependence in these models is generally constrained only from phenomenological considerations.

In this paper we explore the possibility to characterize the rapidity dependence of the initial state directly from high-energy QCD. Based on the phenomenologically successful impact parameter dependent Glasma model (IP-Glasma) -- which in its original form provides boost-invariant initial conditions --  we develop a new initial state model within the Color Glass Condensate (CGC) effective field theory of high-energy QCD \cite{Iancu:2003xm,Gelis:2010nm}, which extends the properties of the initial state to three spatial dimensions.  We obtain the longitudinal rapidity profiles of the collision by consistently including the JIMWLK \cite{Jalilian-Marian:1997xn,Jalilian-Marian:1997jx,Jalilian-Marian:1997gr,Iancu:2000hn,Iancu:2001ad} rapidity evolution for both colliding nuclei up to the measured rapidities. Within this framework, we extract the distributions of produced gluon fields and determine the energy momentum tensor in the transverse plane for discrete values of space-time rapidity, leading to a fully three dimensional initial state which can be employed in hydrodynamic simulations.

Our discussion is organized as follows. In Section \ref{sec:model} we first introduce the 3-D Glasma model detailing in three subsections the calculation of the Wilson lines at the initial rapidity (Section \ref{subsec:init}), the JIMWLK evolution towards smaller $x$ (Section \ref{subsec:jimwlk}), and the computation of the initial state in the forward light-cone for each rapidity (Section \ref{subsec:after}). We present results in Section \ref{sec:results}, separated into the rapidity dependence and fluctuations of the gluon multiplicity (Section \ref{subsec:mult}), and the rapidity dependence of the geometry of the produced system (Section \ref{subsec:geometry}). We conclude and present an outlook in Section \ref{sec:conc}.

\section{3-D Glasma model} \label{sec:model}
Within the Color Glass Condensate effective field theory incoming nuclei are described by color currents
\begin{align}\label{eq:currents}
J_1^{\mu}(x)&=\delta^{\mu+}\rho_{1}^a(x^-,\xt)\, t^a \text{~or~}\notag \\
J_2^{\mu}(x)&=\delta^{\mu-}\rho_{2}^a(x^+,\xt)\, t^a\; ,
\end{align}
(where $t^a$ are the generators of $SU(N_c)$ in the fundamental representation) which determine the gluon fields in the incoming nuclei (1 and 2) that are moving in the $x^+$ and $x^-$ direction, respectively. While in each event the color charges $\rho_{1/2}^a(x^\pm,\xt)$ are distributed randomly inside each nucleus, their statistical properties can be constrained from independent measurements, e.g. in DIS experiments \cite{Kowalski:2003hm,Rezaeian:2012ji}.

When the density of color charges becomes non-perturbatively large at high-energies $\rho \sim 1/g$, it can be shown \cite{Kovner:1995ja,Gelis:2008rw} that the early-time dynamics of the collision is accurately described by the solutions to the classical Yang-Mills equation
\begin{equation}\label{eq:YM1}
[D_{\mu},F^{\mu\nu}] = J^\nu\,,
\end{equation}
to leading order in $\alpha_S$. Before the collision the solution to the classical Yang-Mills equations \,(\ref{eq:YM1}) in Lorentz gauge $(\partial_\mu A^\mu = 0)$ can be immediately inferred as
\begin{equation}
A_{1,2}^\pm (x^\mp, \xt) = -\frac{\rho^{1,2}(x^\mp,\xt)}{\boldsymbol{\nabla}_\perp^2}\,, \quad A^{i}=0\;,
\end{equation}
which in the limit of infinitely thin sheets of color charges $\rho_{1/2} \propto \delta(x^{\mp})$ has support only on the light-cones. However, for practical purposes it is often more convenient to work in Fock-Schwinger gauge $(x^{-}A^{+}+x^{+}A^{-}= 0)$, where the corresponding result is found by a gauge transformation involving the light-like Wilson lines
\begin{equation}\label{eq:wilson}
V_\mathbf{x}^{1,2} = P \exp\left({-ig\int dx^{\mp} \frac{\rho_{1,2}(x^\mp,\xt)}{\boldsymbol{\nabla}_T^2} }\right)\,.
\end{equation}
Outside the forward light-cone the solution of Eq.~(\ref{eq:YM1}) in Fock-Schwinger gauge is then given by pure gauge solutions\,\cite{McLerran:1994ni,*McLerran:1994ka,*McLerran:1994vd,JalilianMarian:1996xn,Kovchegov:1996ty}   
\begin{align}\label{eq:sol}
A^{\pm} = 0\;, \quad  A^i_{1,2}(\xt) &= \theta(x^\mp)\frac{i}{g}V_\mathbf{x}^{1,2}\partial_i V^{\dag\,1,2}_\mathbf{x}\,,  
\end{align}
where $i=1,2$ denotes the transverse spatial Lorentz index. Similarly, the structure of the gauge fields immediately after the collision ($\tau=0^{+}$) can also be obtained analytically and takes the form
\cite{Kovner:1995ja,Kovner:1995ts}:
\begin{align}
  A^i &= A^i_{1} + A^i_{2}\,,\label{eq:init1}\\
  A^\eta &= -\frac{E^\eta}{2} = \frac{ig}{2}\left[A^i_{1},A^i_{2}\right]\,,\label{eq:init2}
\end{align}
associated with boost-invariant flux tubes of longitudinal color-electric and color-magnetic fields \cite{Lappi:2006fp}. 

Starting from the initial conditions at $\tau=0^+$, the early-time dynamics in the forward light-cone ($\tau>0$) can be studied analytically in the small $\tau$ limit \cite{Chen:2015wia} or numerically using lattice gauge theory techniques \cite{Krasnitz:1998ns}. However, it turns out that the boost-invariant nature of the field configurations in Eqs.\,(\ref{eq:init1})-(\ref{eq:init2}) is preserved under the classical Yang-Mills dynamics. Consequently, to explore the properties of the initial state in all three dimensions, one has to consider the effect of higher order corrections to the classical Yang-Mills dynamics which can be of quantitative importance even at the earliest times.

While next-to-leading order (NLO) corrections to the classical Yang-Mills dynamics are essential to understand e.g. the dynamics of the thermalization process at early times \cite{Epelbaum:2013waa,Berges:2013fga,Berges:2014yta,Gelis:2016upa}, a complete understanding of the different effects of NLO corrections is only slowly emerging. However, it was shown \cite{Gelis:2008rw} that for inclusive observables, such as e.g. the energy momentum tensor, an important subset of the next-to-leading order corrections which are enhanced by a large rapidity separation $\Delta y_{1/2}$, where $\Delta y_{1/2}$ denotes the rapidity difference between the incoming nucleus (1 or 2) and the rapidity $y_{\rm{obs}}$ where the measurement is performed, can be resummed to all orders and absorbed into a re-definition of the Wilson lines. While any inclusive observable at a given rapidity $y_{\rm{obs}}$ is still computed by solving the classical Yang-Mills equations, the Wilson lines  $V_\mathbf{x}^{1,2}$ become explicitly dependent on the respective rapidity scale $\Delta y_{1/2}$, with the rapidity evolution described by the JIMWLK renormalization group equation \cite{Jalilian-Marian:1997xn,Jalilian-Marian:1997jx,Jalilian-Marian:1997gr,Iancu:2000hn,Iancu:2001ad}.

\begin{figure}[tb]
\includegraphics[width=0.48\textwidth]{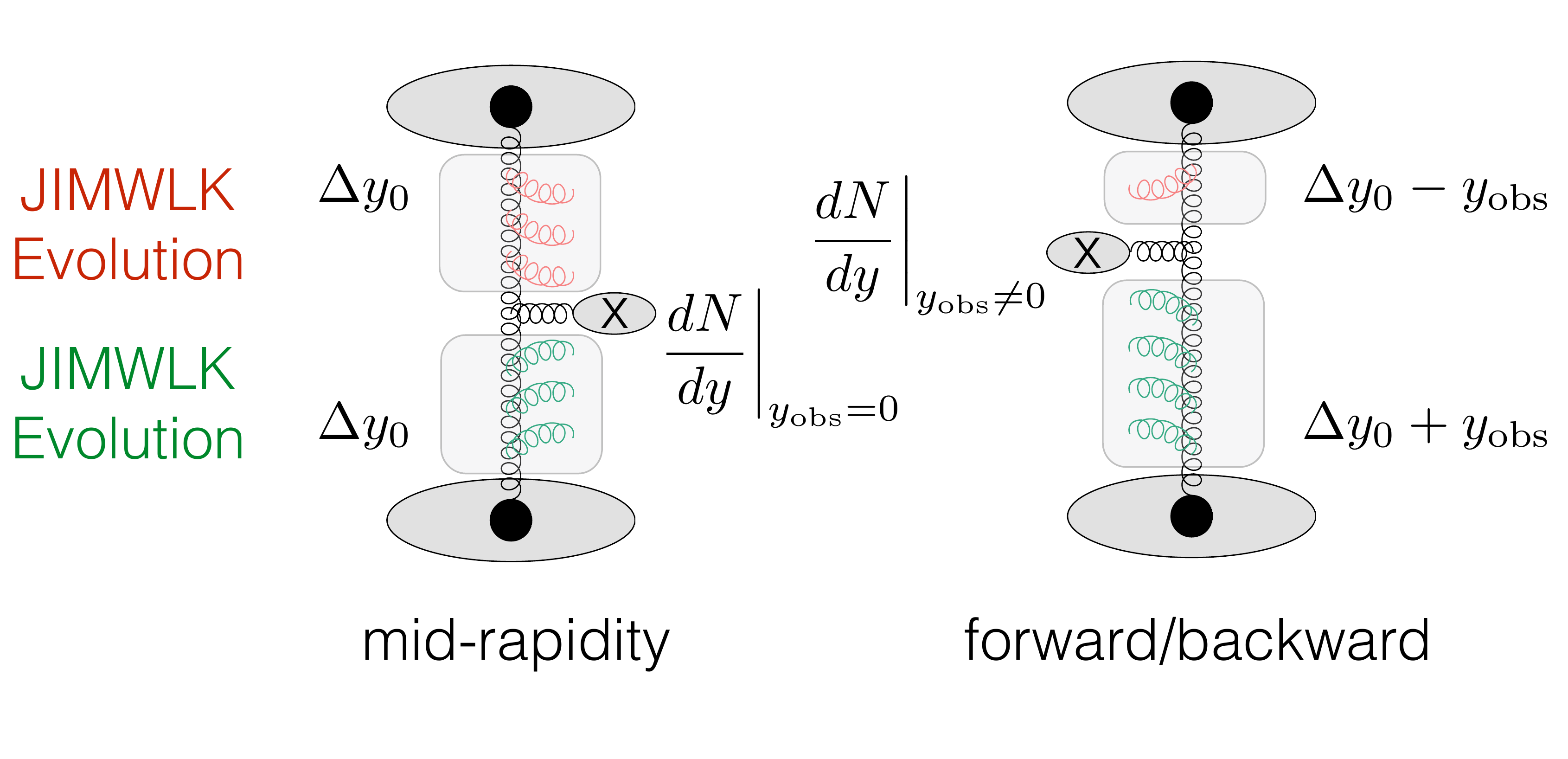}
\caption{
\label{fig:JIMWLKScheme} Computation scheme for single inclusive observables (left) at mid-rapidity and (right) forward/backward rapidity. Starting from the IP-Glasma parametrization of gluon distributions at an initial rapidity scale, JIMWLK rapidity evolution is applied to both nuclei to resum leading logarithmic corrections to inclusive observables. While the same amount of rapidity evolution $\Delta y_{1/2}=\Delta y_{0}$ is applied to both nuclei to compute observables at mid-rapidity ($y_{{\rm obs}}=0$), the amount of evolution in the two nuclei is different $\Delta y_{1/2}=\Delta y_{0}\pm y_{{\rm obs}}$ when computing observables at forward/backward rapidities  ($y_{{\rm obs}}\neq0$).}
\end{figure}

A compact summary of the computational scheme to compute single inclusive observables at leading logarithmic accuracy is depicted in Fig.\,\ref{fig:JIMWLKScheme}. Starting from a parametrization of the distributions of color charges at an initial rapidity scale $y_0$ in the IP-Glasma model \cite{Schenke:2012wb,Schenke:2012hg}, evolution towards larger rapidity separations $\Delta y_{0}\pm y_{\rm{obs}}$ is then performed by numerically solving the JIMWLK equations \cite{Jalilian-Marian:1997xn,Jalilian-Marian:1997jx,Jalilian-Marian:1997gr,Iancu:2000hn,Iancu:2001ad}. Based on the evolved Wilson line configurations one constructs the classical Yang-Mills fields in the forward light-cone at any given rapidity $y_{\rm{obs}}$ according to Eqs.~(\ref{eq:init1})-(\ref{eq:init2}). Subsequently, the classical Yang-Mills equations of motion are solved numerically in the forward light-cone to compute the observables of interest. 

As we pointed out, the above framework of high-energy factorization has been rigorously derived for single-inclusive observables at different rapidities \cite{Gelis:2008rw}, and shown to hold also for inclusive multi-gluon production, as long as the rapidity separation between the particles is maximally $\sim \alpha_s^{-1}$ \cite{Gelis:2008ad,Gelis:2008sz}. This framework has been the basis of CGC based calculations of the ``ridge effect'' in heavy ion \cite{Dusling:2009ni,Lappi:2009xa}, p+p \cite{Dumitru:2010iy,Dusling:2012iga}, and p+A \cite{Dusling:2012cg,Dusling:2012wy,Dusling:2013qoz,Schenke:2015aqa} collisions.

However, some cautionary remarks are in order regarding its applicability to the typical observables of interest for the phenomenological description of heavy-ion collisions. The event-by-event hydrodynamic description of heavy-ion collisions requires the knowledge of unequal rapidity correlations over a wide range in rapidity. For example, typical observables require the energy-momentum-tensor to be known simultaneously at two different rapidities, thus one expects corrections to results obtained within the framework of Fig.\,\ref{fig:JIMWLKScheme} as soon as the rapidity separation exceeds $\alpha_s^{-1}$. 

For dilute-dense systems, extended evolution equations have been derived to incorporate corrections to above factorization scheme \cite{Iancu:2013uva}. However, a comprehensive framework to incorporate these effects in nucleus-nucleus collisions is yet to be developed and we intend to return to this issue in a future publication. On the other hand, it is conceivable that the extension of the above factorization framework to a larger class of observables will at least capture the most important effects correctly. We will therefore employ this prescription in the following as a first important step towards constraining the longitudinal profiles of high-energy collisions from first principles.

Details of the implementation of this 3D-Glasma model are described below.

\subsection{Wilson lines at the initial rapidity scale}\label{subsec:init}
We now discuss how to numerically determine the gluon fields in the two incoming nuclei at the initial rapidity scale. First, we need to determine the color charge densities that appear in Eq.\,(\ref{eq:currents}). To do so we follow previous works \cite{Schenke:2012wb,Schenke:2012hg} and use the impact parameter dependent saturation (IPSat) model \cite{Bartels:2002cj,Kowalski:2003hm} as the starting point. The model provides a parametrization of the saturation scale $Q_s$ as a function of Bjorken $x$ and transverse position $\xt$, which is constrained from deep inelastic scattering (DIS) data \cite{Rezaeian:2012ji}. 

Using the thickness function of the nucleus, obtained by sampling individual nucleons from a Woods-Saxon distribution and assigning a Gaussian thickness function to each nucleon \cite{Schenke:2012wb,Schenke:2012hg}, the IPSat model provides $Q_s(\xt,x)$ at any given transverse position $\xt$ and the Bjorken $x$ that corresponds to the initial rapidity at a given collision energy $\sqrt{s}$.  
The color charge density distribution $g^2 \mu(\xt,x)$ is proportional to $Q_s(\xt,x)$, which can fluctuate in each nucleon following \cite{McLerran:2015qxa}.\footnote{While we do not expect such fluctuations to be particularly important for heavy ion collisions, they are required in p+A and p+p collisions to describe the multiplicity distributions \cite{Schenke:2013dpa}.}
\begin{figure*}[tb]
\includegraphics[width=0.98\textwidth]{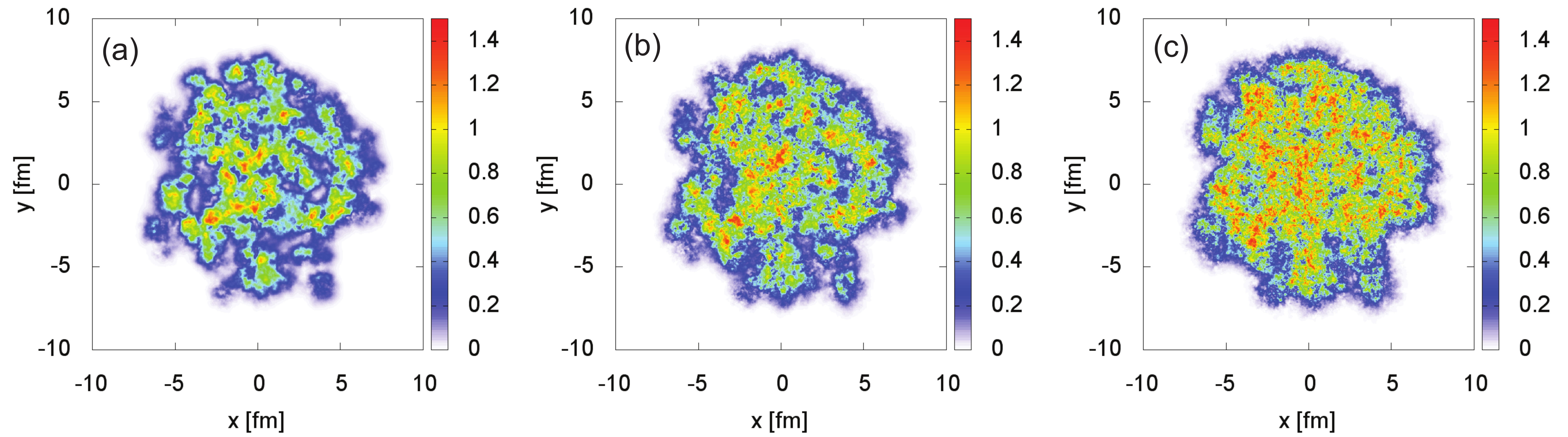}
\caption{
JIMWLK evolution of the gluon fields in one nucleus for $m=0.4\,{\rm GeV}$ and $\alpha_s=0.3$. Shown is $1-{\rm Re}[{\rm tr}(V_{\mathbf{x}})]/N_c$ in the transverse plane at rapidities $Y=-2.4$ ($x\approx 2\times 10^{-3}$) (a), $Y=0$ ($x\approx 2\times 10^{-4}$) (b), and $Y=2.4$ ($x\approx 1.6\times 10^{-5}$) (c) to illustrate the change of the typical transverse length scale with decreasing $x$. The global geometry clearly remains correlated over the entire range in rapidity.  \label{fig:JIMWLK}
}
\end{figure*}
We obtain the Wilson lines at the initial rapidity scale by discretizing the path ordered exponential in Eq.~(\ref{eq:wilson}) according to \cite{Lappi:2007ku}
\begin{equation}\label{eq:wilsonNum}
  V_\mathbf{x} = \prod_{k=1}^{N_y}\exp\left(-ig\frac{\rho_k(\xt)}{\boldsymbol{\nabla}_\perp^2+m^2}\right)\,,
\end{equation}
where we introduce the infrared regulator $m\sim\Lambda_{\rm QCD}$ to suppress Coulomb tails. Color charge distributions $\rho_k$ are sampled from a Gaussian distribution with width $g \mu(\xt,x)$
 \begin{equation}
  \langle \rho_k^a(\xt)\rho_l^b(\yt)\rangle = \delta^{ab}\delta^{kl}\delta^2(\xt-\yt)\frac{g^2\mu^2(\xt)}{N_y}\,,
 \end{equation}
where the indices $k,l=1,2,\dots,N_y$ represent a discretized $x^-$ (or $x^+$) coordinate, and we use $N_y=50$ throughout this work .

\subsection{JIMWLK evolution}\label{subsec:jimwlk}
Starting from the Wilson line configurations at the initial rapidity scale, we perform the JIMWLK renormalization group evolution \cite{Jalilian-Marian:1997xn,Jalilian-Marian:1997jx,Jalilian-Marian:1997gr,Iancu:2000hn,Iancu:2001ad} to the largest rapidity separation of interest as depicted in Fig. \ref{fig:JIMWLKScheme}. 
For our numerical study it is particularly useful to express the JIMWLK hierarchy in terms of a functional Langevin equation for the Wilson lines \cite{Weigert:2000gi,Blaizot:2002xy}.
In \cite{Lappi:2012vw} the following simple form of the Langevin step was derived:
\begin{align}\label{eq:newstep}
  V_{\mathbf{x}}(Y+dY) = \exp \Big\{-i \frac{\sqrt{\alpha_s dY}}{\pi} \int_{\mathbf{z}} K_{\mathbf{x}-\mathbf{z}}\cdot (V_{\mathbf{z}}\boldsymbol{\xi}_{\mathbf{z}}
V^\dag_{\mathbf{z}})\Big\}\notag\\ \times V_{\mathbf{x}}(Y) \exp \Big\{ i \frac{\sqrt{\alpha_s dY}}{\pi} \int_{\mathbf{z}} K_{\mathbf{x}-\mathbf{z}}\cdot \boldsymbol{\xi}_{\mathbf{z}}\Big\}\,,
\end{align}
where the noise $\boldsymbol{\xi}_{\mathbf{z}} = (\xi_{\mathbf{z},1}^{a} t^a,\xi_{\mathbf{z},2}^{a} t^a)$ is taken to be Gaussian and local in transverse coordinate, color, and rapidity: $\langle \xi_{\mathbf{z}, i}^b(Y)\rangle = 0$ and
\begin{equation}\label{eq:noise}
  \langle \xi_{\mathbf{x}, i}^a(Y) \xi_{\mathbf{y}, j}^b(Y')\rangle = \delta^{ab}\delta^{ij}\delta_{ \mathbf{x}\mathbf{y} }^{(2)} \delta(Y-Y')\,,
\end{equation}
and the perturbative JIMWLK kernel is given by
\begin{equation}\label{eq:K}
  K_{\mathbf{x}-\mathbf{z}} = \frac{(\mathbf{x}-\mathbf{z})}{(\mathbf{x}-\mathbf{z})^2}\,.
\end{equation}

As discussed in \cite{Schlichting:2014ipa}, the perturbative kernel needs to be regularized at large distance scales to limit growth in impact parameter space.
This is done using the modified kernel
\begin{equation}\label{eq:modKernel}
  K^{\text{(mod)}}_{\mathbf{x}-\mathbf{z}} = m|\mathbf{x}-\mathbf{z}|~K_{1}(m |\mathbf{x}-\mathbf{z}|)~K_{\mathbf{x}-\mathbf{z}}.
\end{equation}
Here $K_{1}(x)$ is the modified Bessel function of the second kind. At small arguments $x K_{1}(x)=1+\mathcal{O}(x^2)$, such that the kernel is unmodified, while for large arguments $K_{1}(x)=\sqrt{\frac{\pi}{2x}} e^{-x}$ decays exponentially with the infrared regulator $m$ chosen to be equal to that introduced in Eq.\,(\ref{eq:wilsonNum}).
On the level of the Balitsky-Kovchegov equation \cite{Balitsky:1995ub,Kovchegov:1999yj}, which is the large $N_c$-limit of the JIMWLK equation, it is known that next-to leading order effects \cite{Balitsky:2008zza} slow down the evolution \cite{Iancu:2015vea,Iancu:2015joa,Lappi:2016fmu}. Here, we adjust the evolution speed by treating the strong coupling constant $\alpha_s$ as a free parameter.

In practice, we need to determine the Wilson lines of a nucleus at all $x$ relevant for a certain collision energy and rapidity range. To do so we first compute the Wilson lines at the largest $x$ value using the IPSat/IP-Glasma model. For a left moving nucleus at LHC energies of $\sqrt{s} = 2.76\,{\rm TeV}$ at $y=-2.4$, which is the largest (negative) rapidity we consider, this is $x=(\langle p_T \rangle/\sqrt{s}) \exp(2.4) \approx 0.002$ for the typical $\langle p_T\rangle \approx\,0.5\,{\rm GeV}$. We then solve the JIMWLK equation (\ref{eq:newstep}) up to rapidity $y=2.4$, which corresponds to $x\approx 1.6 \times 10^{-5}$. For the right moving nucleus the $x$ values are simply reversed.

In Fig.\,\ref{fig:JIMWLK} we show the transverse structure of one nucleus employing the quantity $1-{\rm Re}[{\rm tr}(V_{\mathbf{x}})]/N_c$ at different values of $x$ using $m=0.4\,{\rm GeV}$ and $\alpha_s=0.3$. As previously demonstrated in \cite{Dumitru:2011vk,Schlichting:2014ipa} the increase of the saturation scale $Q_s$ with decreasing $x$ leads to a reduction of the characteristic transverse length scale $\sim 1/Q_s$, which is clearly visible in Fig.\,\ref{fig:JIMWLK}. Despite significant changes on smaller scales, one also observes that the large scale structure of the nucleus is only mildly modified even after evolution of several units in rapidity.
\begin{figure*}[tb]
\includegraphics[width=0.9\textwidth]{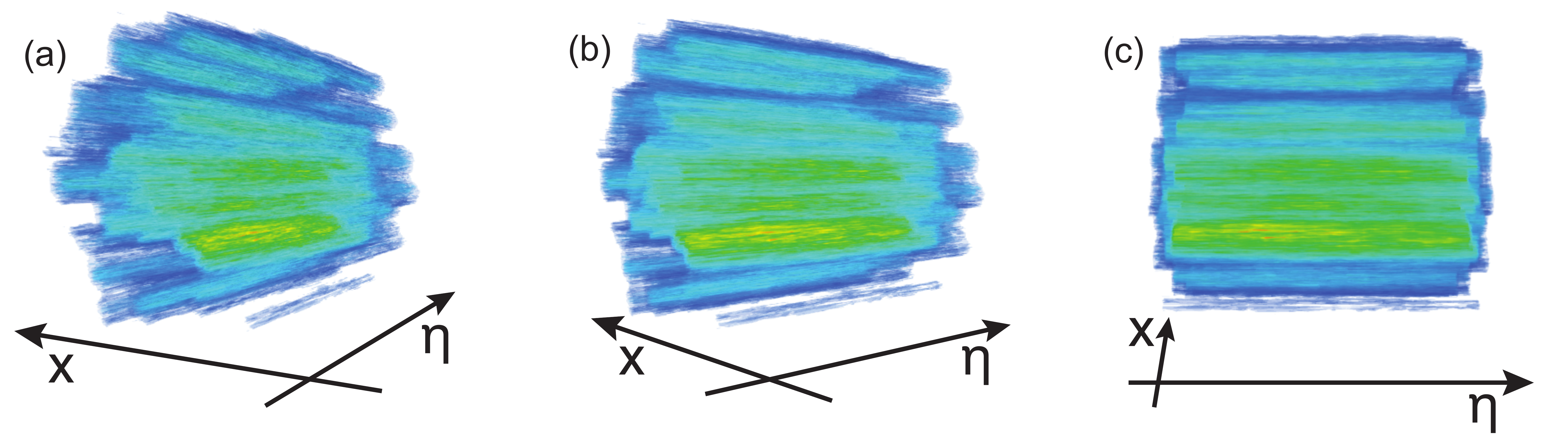}
\caption{
View of the three dimensional distribution of $T^{\tau\tau}$ in a single event from different angles, covering the entire transverse plane and 4.8 units in rapidity.\label{fig:3dimage}}
\end{figure*}

\subsection{Gluon fields after the collision}\label{subsec:after}
Using the Wilson lines of the incoming nuclei at rapidities $\Delta Y_{1/2} = \Delta Y_0 \pm Y_{\rm obs}$, the gluon fields in the future light-cone are determined numerically for each slice in rapidity $y_{\rm obs}$. Based on the procedure outlined in \cite{Krasnitz:1998ns} we determine the discretized analogue of Eqs.\,(\ref{eq:init1})-(\ref{eq:init2}).
Given these initial conditions, the source free Yang-Mills equations are solved forward in time, for each rapidity separately (see e.g. \cite{Schenke:2012hg}). From the resulting fields at $\tau=0.2\,{\rm fm}/c$ we compute the gluon multiplicity and the energy momentum tensor $T^{\mu\nu}$. 

The gluon distribution can be obtained by measuring equal-time correlation functions of the gauge fields after imposing Coulomb gauge $\left.\partial_{i}A^{i}\right|_{\tau}=0$ at a given time \cite{Berges:2013fga,Berges:2013eia}:
\begin{eqnarray}\label{eq:singleParticleDist}
\left.\frac{dN}{d^{2}\kt dy}\right|_{\tau}=\frac{1}{(2\pi)^2} \sum_{\lambda,a} \left| \tau g^{\mu\nu} \Big( \xi_{\mu}^{\lambda,\kt*}(\tau) \overleftrightarrow{\partial_{\tau}} A_{\nu}^{a}(\tau,\kt)\Big) \right|^2 \nonumber \\
\end{eqnarray}
where  $g^{\mu\nu}=(1,-1,-1,-\tau^{-2})$ denotes the Bjorken metric and $\lambda=1,2$ labels the two transverse polarizations. In Coulomb gauge the mode functions take the form
\begin{eqnarray}
\xi_{\mu}^{(1),\kt}(\tau)&=& \frac{\sqrt{\pi}}{2|\kt|}  \begin{pmatrix} -k_y \\ k_x \\ 0 \end{pmatrix}  H^{(2)}_{0}(|\kt|\tau) \;, \label{eq:A1} \\
\xi_{\mu}^{(2),\kt}(\tau)&=&\frac{\sqrt{\pi}}{2|\kt|}  \begin{pmatrix} 0 \\ 0 \\  k_T\tau \end{pmatrix}  H^{(2)}_{1}(|\kt|\tau) \;, \label{eq:A2}
\end{eqnarray}
where $H^{(2)}_{\alpha}$ denote the Hankel functions of the second type and order $\alpha$ (see \cite{Berges:2013fga} for details).

\section{Results}\label{sec:results}
For illustration we first present 3-D plots of $T^{\tau\tau}$ in a single event at $\sqrt{s}=2.76\,{\rm TeV}$ in Fig.\,\ref{fig:3dimage}. We find that in each event the distribution of deposited energy is dominated by approximately boost invariant flux tube like structures, with a typical transverse size of a nucleon. Deviations from boost invariance introduced by the JIMWLK evolution of both nuclei are clearly visible as well. Variations on both small and large transverse and longitudinal distance scales can be observed. 

These structures in rapidity are quantified in the following, where we compute the rapidity dependence of the gluon multiplicity and its fluctuations as well as the rapidity variation in value and orientation of eccentricities. These quantities should have a direct effect on various experimental observables, such as the charged particle rapidity spectra, the pseudo-rapidity dependent factorization ratio $r_n(\eta_1,\eta_2)$ \cite{Khachatryan:2015oea}, and Legendre coefficients of two-particle multiplicity \cite{Bzdak:2012tp} and eccentricity correlations in rapidity.

\subsection{Rapidity dependence of the multiplicity}\label{subsec:mult}
\begin{figure}[H]
\includegraphics[scale=0.8]{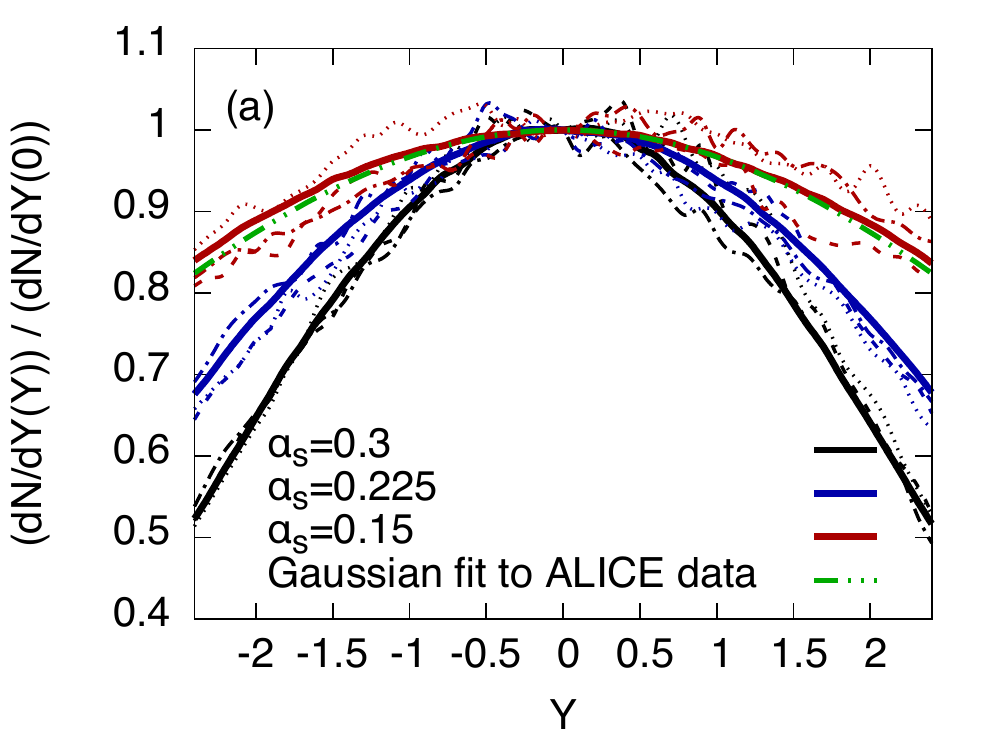}
\includegraphics[scale=0.8]{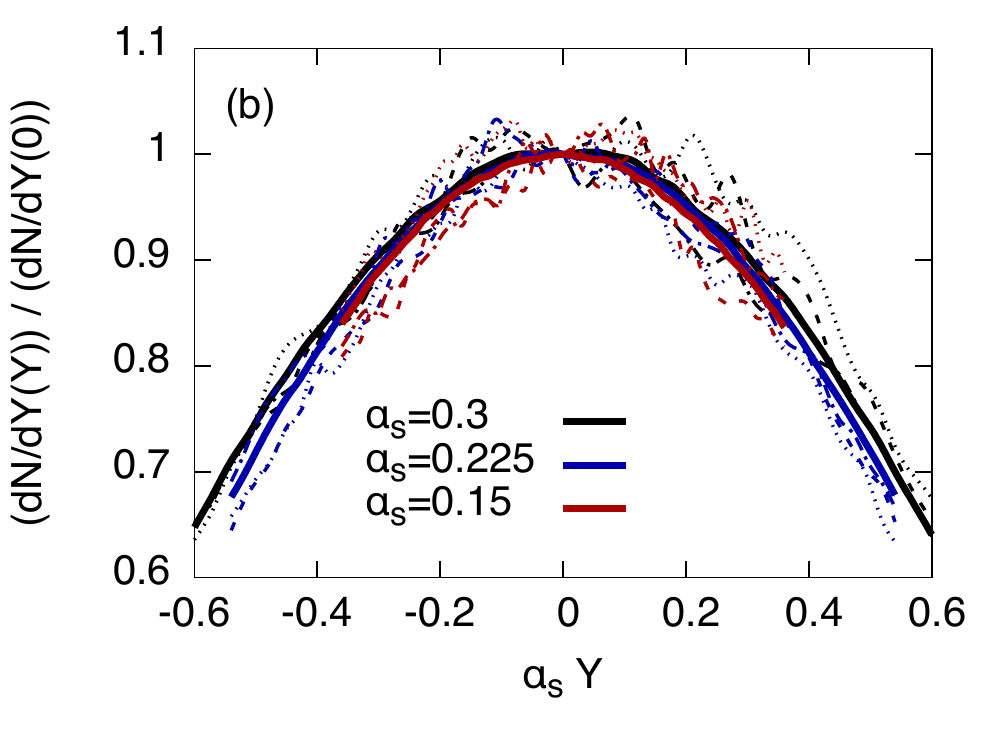}
\caption{\label{fig:mult} a) Gluon multiplicity relative to its value at $Y=0$ for $\alpha_s=0.15$, $\alpha_s=0.225$, and $\alpha_s=0.3$ using $m=0.4\,{\rm GeV}$. The various dashed lines show results from three single events for each value of the coupling constant. The green dash-dot-dotted line is a Gaussian fit to the charged hadron $dN_{\rm ch}/dY$ data from ALICE \cite{Abbas:2013bpa}. b) The same results plotted vs. $\alpha_s Y$. }
\end{figure}

We first present results for the rapidity dependence of the transverse momentum integrated gluon multiplicity.
Fig.~\ref{fig:mult} shows the event averaged gluon multiplicity relative to its value at $Y=0$ for $\alpha_s=0.15$, $\alpha_s=0.225$, and $\alpha_s=0.3$ and $m=0.4~{\rm GeV}$. The dependence on the coupling $\alpha_s$ is clearly visible. In particular, we find approximate scaling with $\alpha_s Y$, as demonstrated explicitly in the lower panel of the figure. The statistical errors are smaller than the width of the line. To demonstrate the event-by-event fluctuations, we also show results from three single events using thin lines. To get a sense of the magnitude of the rapidity dependence we compare to a Gaussian fit (width $\sigma = 3.86$) of experimental data for $dN_{\rm ch}/dY$ from ALICE, also scaled by the value at $Y=0$ \cite{Abbas:2013bpa}. Hydrodynamic evolution will broaden the initial distribution in space-time rapidity to produce somewhat broader $dN_{\rm ch}/dY$ spectra (see e.g. \cite{Nonaka:2006yn}). We thus conclude that when characterizing the evolution speed by a constant $\alpha_s$, it needs to be 0.15 or greater to generate results compatible with the experimental data.  
In order to compare to evolution speeds quoted in the description of structure functions we compute 
\begin{equation}
  \lambda = \frac{d \ln Q_s^2}{dY}\,.
\end{equation}
$Q_s$ is defined as the inverse of $r$ at which the dipole amplitude $\mathcal{N} = {\rm tr}\, \langle1- V^\dag(\mathbf{b}+\mathbf{r}/2) V(\mathbf{b}-\mathbf{r}/2)\rangle/N_c$, where the average is over $\mathbf{b}$, reaches the value 0.15. \footnote{We constrain ourselves to small values of the dipole amplitude because at large $r$ non-perturbative effects that are not included in our prescription affect its value. \cite{Schlichting:2014ipa}}
We further neglect the detailed $Y$ dependence of $\lambda$ and quote a range of $\lambda$ values over the considered $Y$ range.
We find $\lambda \approx 0.28-0.3$ for $\alpha_s=0.15$ and $\lambda\approx 0.6-0.8$ for $\alpha_s=0.3$. Values of $\lambda=0.2-0.3$ are consistent with experimental data on structure functions \cite{Golec-Biernat:1998js,Iancu:2003ge,Rezaeian:2013tka}. 

We note that the results presented in Fig.\,\ref{fig:mult} depend only mildly on the infrared regulator $m$. For $m = 0.2\,{\rm GeV}$ we find a slighty faster change with $Y$, leading to an approximately $15\%$ smaller $dN/dY$ at $Y = \pm2.4$.

The single event rapidity distributions $dN/dY$ are then used to determine the two-particle rapidity correlation function \cite{Vechernin:2013vpa}
\begin{equation}
  C(Y_1,Y_2) = \frac{\langle N(Y_1)N(Y_2) \rangle}{\langle N(Y_1) \rangle \langle N(Y_2) \rangle}\,,
\end{equation}
where $N(y)=dN/dy$.
\begin{figure}[tb]
 \begin{minipage}{\linewidth}
      \centering
       \includegraphics[width=10cm]{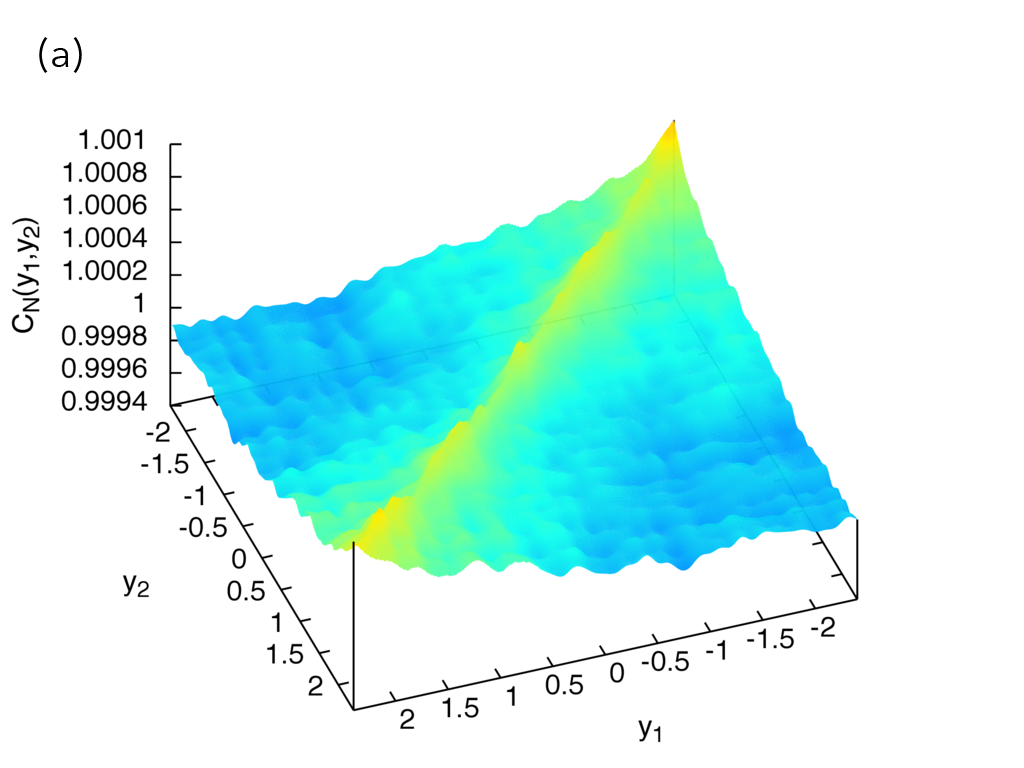}
       \includegraphics[width=10cm]{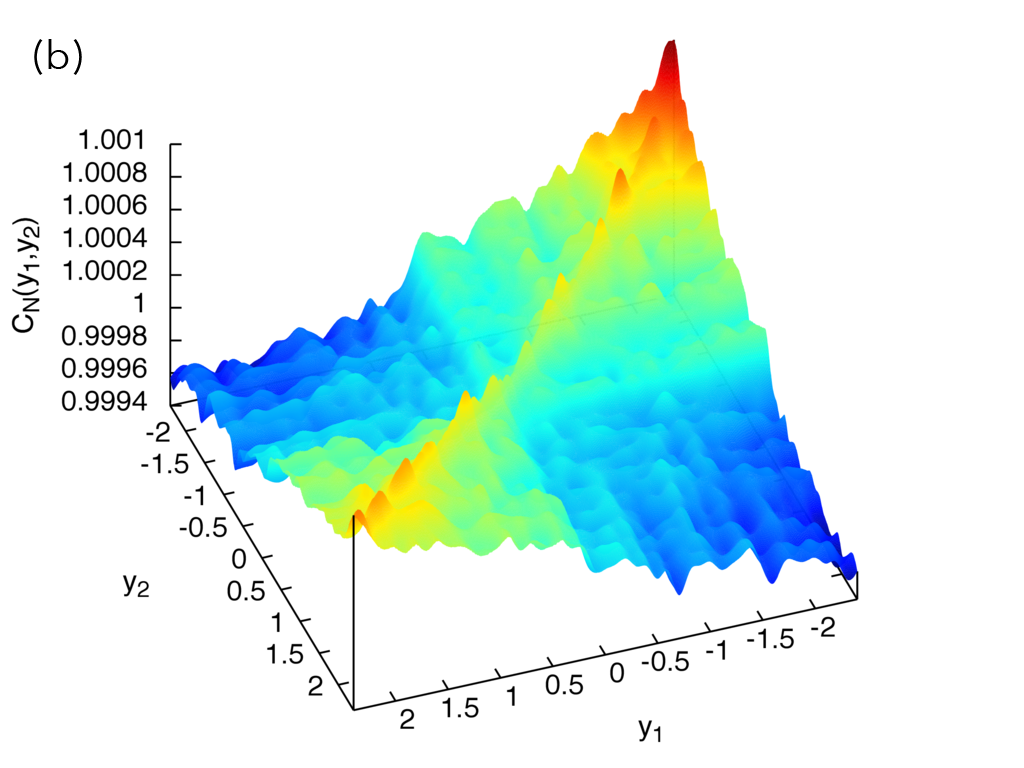}
\end{minipage}
\caption{ The correlation function $C_N(Y_1,Y_2)$ for $m=0.4\,{\rm GeV}$ and $\alpha_s=0.15$ (a) and $\alpha_s=0.3$ (b). \label{fig:CN}}
\end{figure}
In Fig.\,\ref{fig:CN} we show the result for the normalized correlation function 
\begin{equation}\label{eq:cn}
  C_N(Y_1,Y_2) = \frac{C(Y_1,Y_2)}{ C_p(Y_1) C_p(Y_2)}\,,
\end{equation}
where the normalization factors $C_p(Y_{1}) = \frac{1}{2Y}\int_{-Y}^{Y} C(Y_1,Y_2) dY_{2}$ and  $C_p(Y_{2}) = \frac{1}{2Y}\int_{-Y}^{Y} C(Y_1,Y_2) dY_{1}$ are chosen such that $C_N(Y_1,Y_2)$ is normalized to be one on average. The general structure and magnitude of the correlation function is similar to that observed by the ATLAS Collaboration \cite{ATLAS-CONF-2015-020}. The expected stronger rapidity dependence for the larger $\alpha_s=0.3$ compared to $\alpha_s=0.15$ is clearly visible.

In analogy to the experimental data \cite{ATLAS-CONF-2015-020}, our result for the gluon $C_N$ can be expanded in Legendre polynomials.
Following \cite{Bzdak:2012tp,Jia:2015jga} the Legendre coefficients are given by
\begin{align}\label{eq:legendre}
  a_{n,m} =& \int C_N(Y_1,Y_2) \nonumber\\ &~~ \times \frac{T_n(Y_1) T_m(Y_2) + T_n(Y_2) T_m(Y_1)}{2} \frac{dY_1}{Y} \frac{dY_2}{Y}\,,
\end{align}
where $T_n(Y_p) = \sqrt{n+1/2}\,P_n(Y_p/Y)$ and $P_n$ are the standard Legendre polynomials.
The $a_{n,m}$ are related to the Legendre coefficients of the single particle distribution $a_n$ via
$a_{n,m}=\langle a_n a_m\rangle$, where the $a_n$ are defined through $N(y)/\langle N(y) \rangle = 1 + \sum_n a_n T_n(y)$ \cite{ATLAS-CONF-2015-020,ATLAS-CONF-2015-051}.

\begin{figure}[tb]
  \includegraphics[width=8cm]{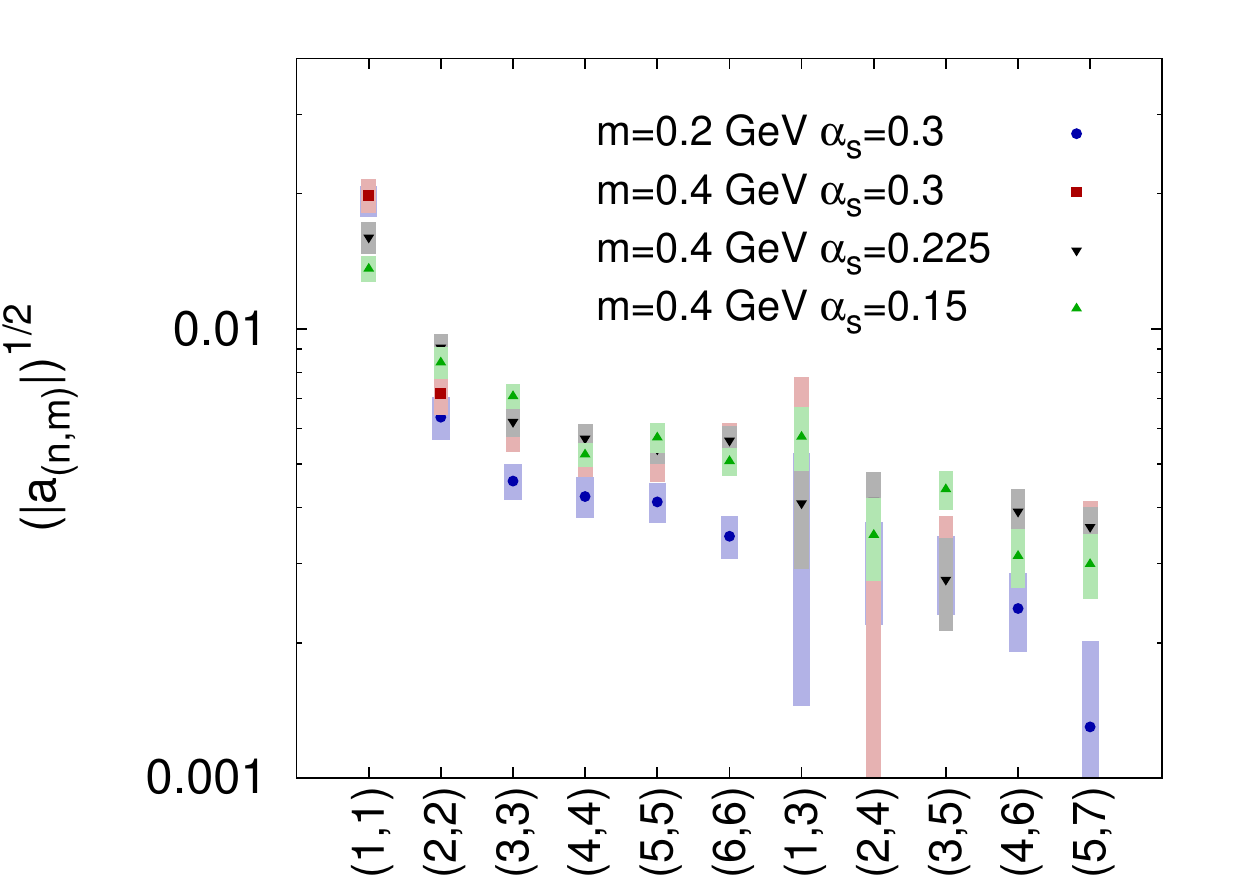}
\caption{  Legendre coefficients $\sqrt{|a_{n,m}|}$ labeled by $(n,m)$ of the gluon multiplicity correlator for different values of $\alpha_s$ and $m$. \label{fig:anm}}
\end{figure}

The results for $\sqrt{|a_{n,m}|}$ with different $\alpha_s$ and mass parameters $m$ are shown in Fig.\,\ref{fig:anm}. 
In the experimental data the coefficient $a_{1,1}$ is the only one insensitive to short range correlations, such as those resulting from resonance decays.
It is further largely unaffected by final state interactions \cite{Monnai:2015sca} and Fig.\,\ref{fig:anm} shows that it is also insensitive to $m$, which makes it a rather robust observable to constrain the evolution speed characterized by $\alpha_s$.
Extrapolating the experimental results where short range correlations are removed and $p_T>0.2\,{\rm GeV}$ to very central events, which we consider here, ATLAS finds $\sqrt{a_{1,1}}\approx 0.015$ \cite{ATLAS-CONF-2015-051}, which is approximately in between our results using $\alpha_s=0.15$ and $\alpha_s=0.225$. We show higher Legendre coefficients for completeness. They can be used for comparison with different initial state models but should not be compared to experimental data, because they are affected both by final state interactions, like hydrodynamic evolution, and short range correlations like those from resonance decays \cite{Monnai:2015sca,Bozek:2015tca}. We note that in p+p collisions $a_{1,1}$ was determined in a model including fluctuations of the saturation scale and evolution of this scale with rapidity in \cite{Bzdak:2015eii}. This should capture important features of our more detailed calculation. 

\begin{figure*}[tb]
 \begin{minipage}{\linewidth}
      \centering
      \begin{minipage}{0.45\linewidth}
              \includegraphics[width=\linewidth]{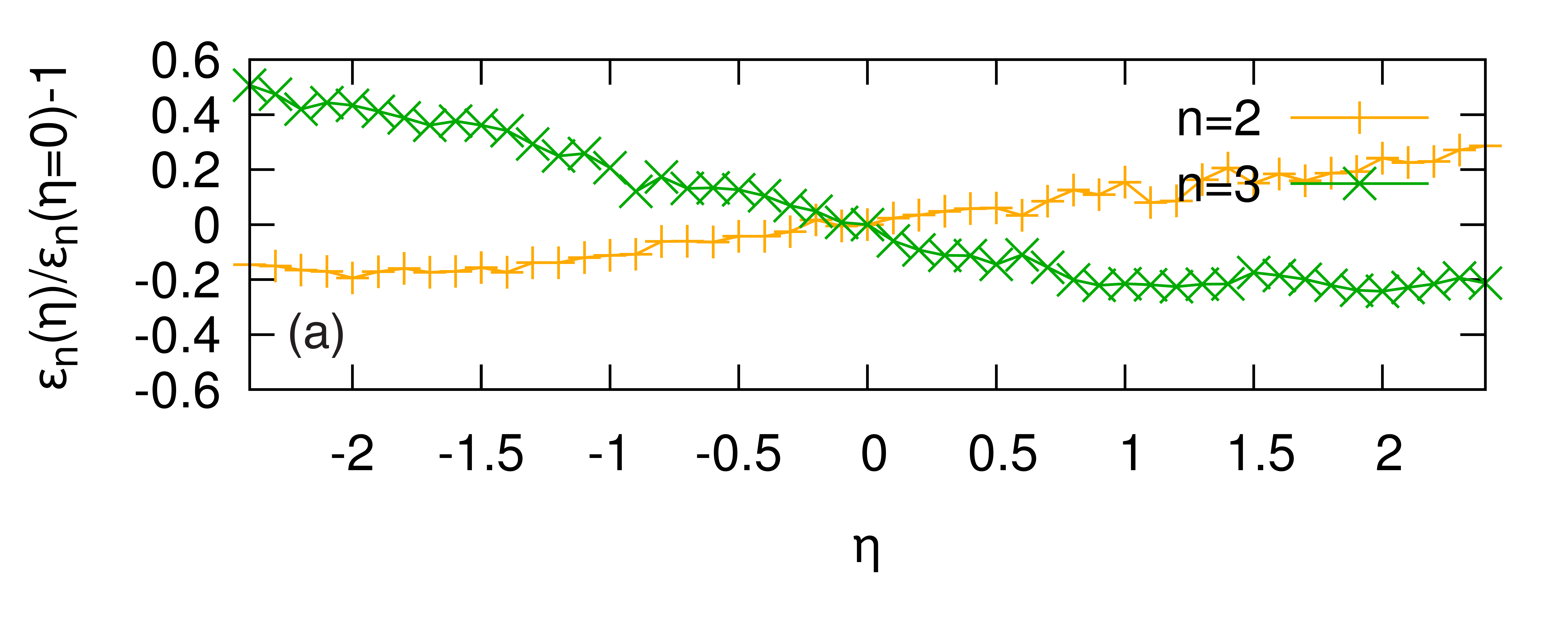} \includegraphics[width=\linewidth]{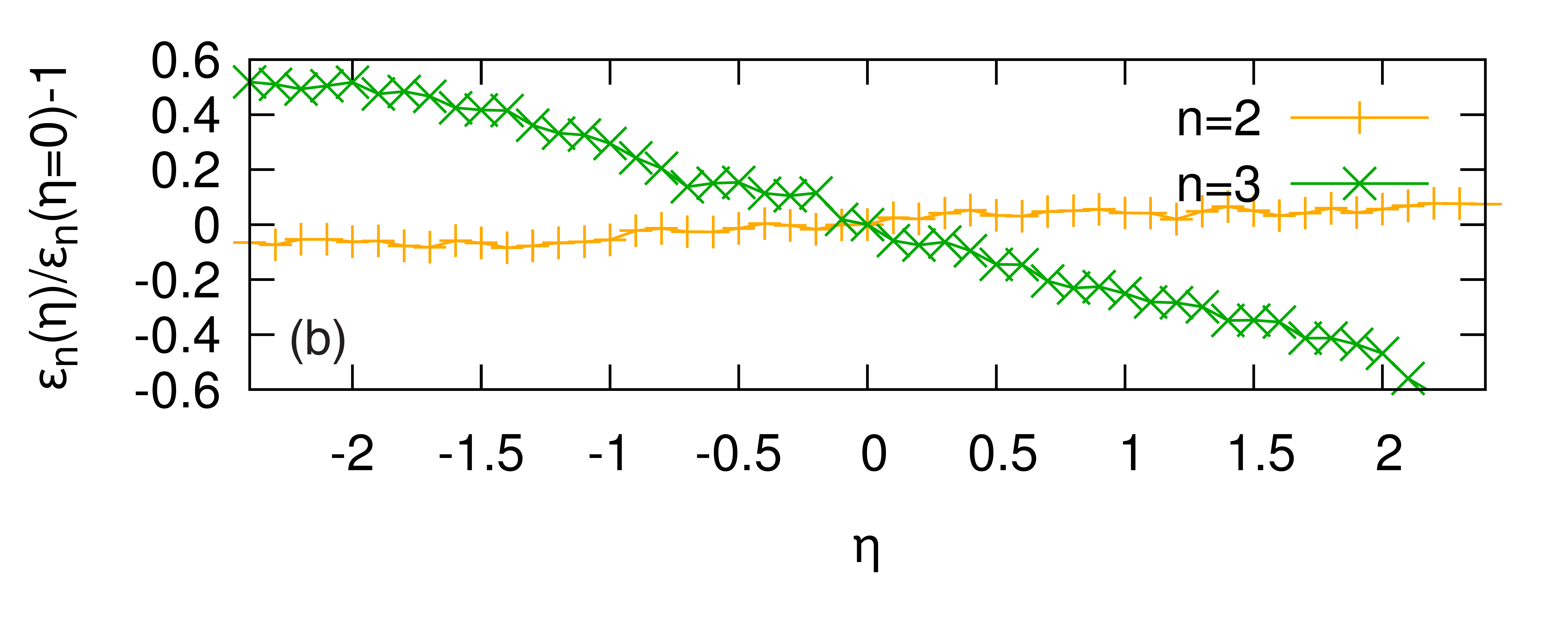} \includegraphics[width=\linewidth]{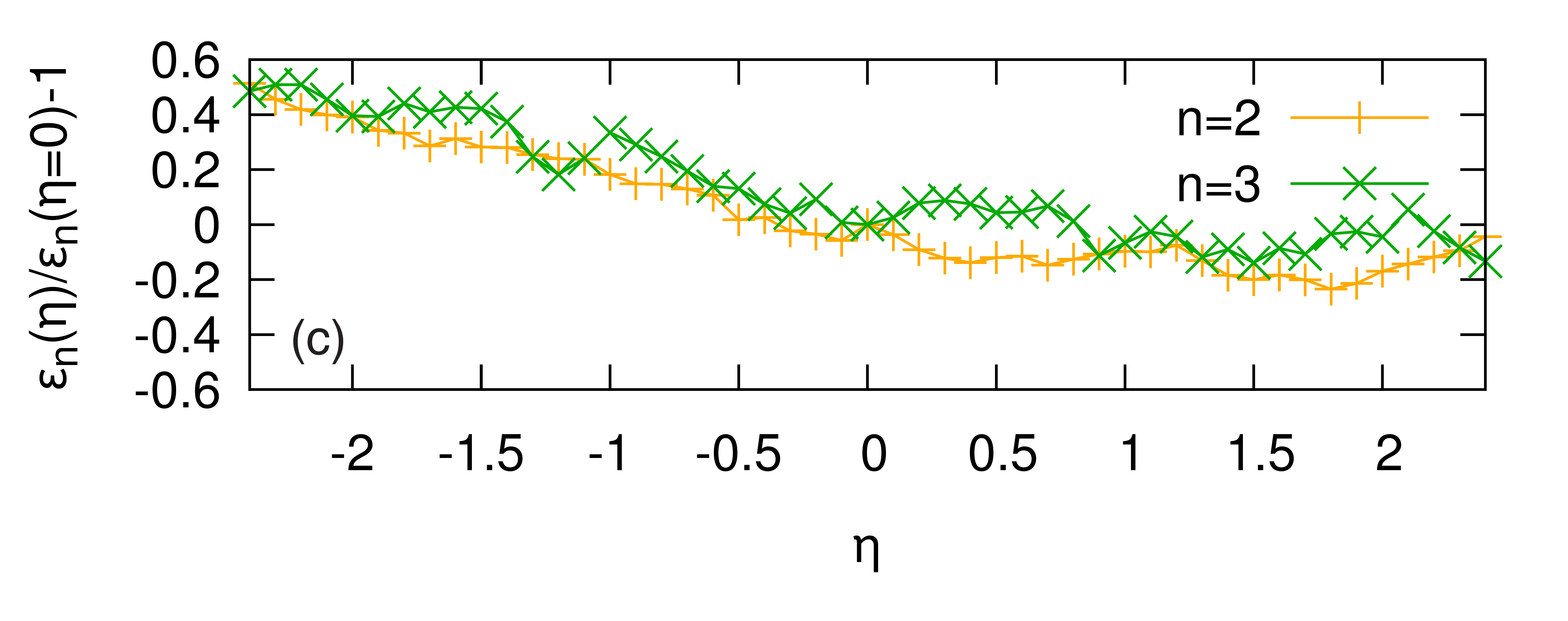}
      \end{minipage}
      \hspace{0.05\linewidth}
      \begin{minipage}{0.45\linewidth}
              \includegraphics[width=\linewidth]{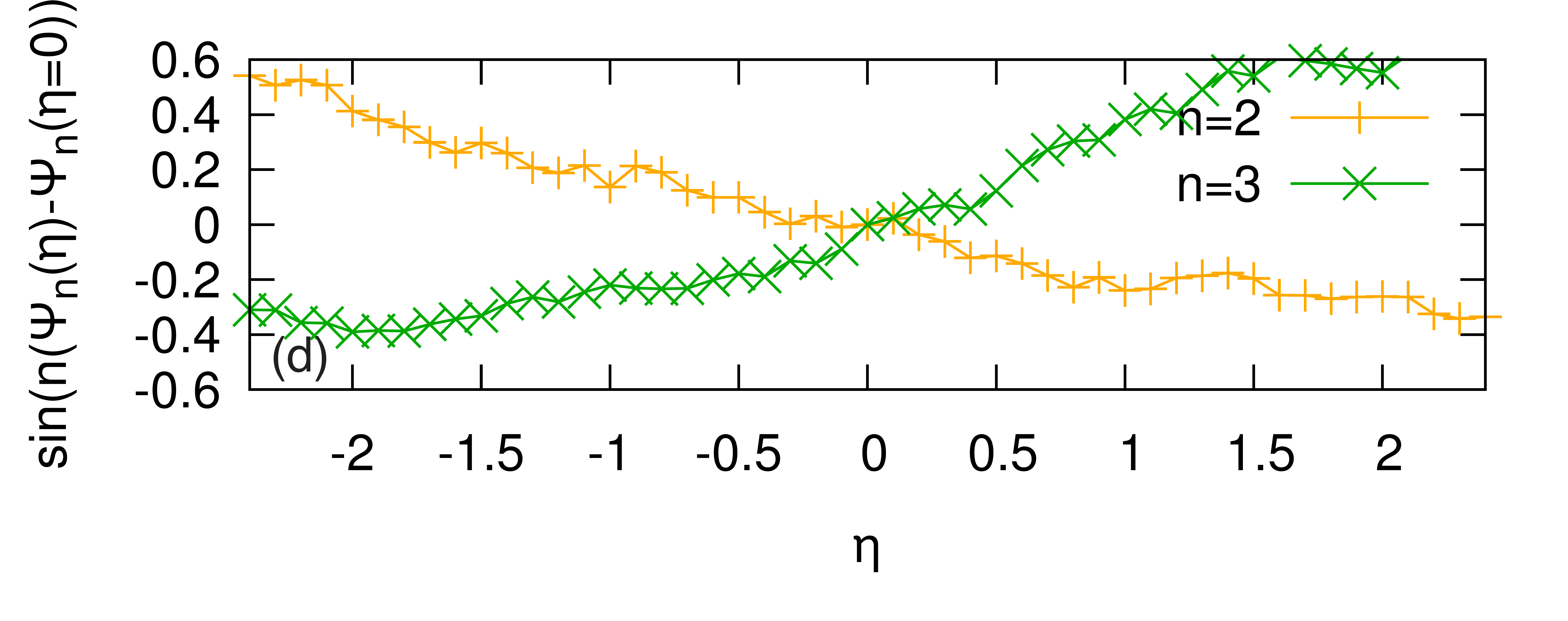}  \includegraphics[width=\linewidth]{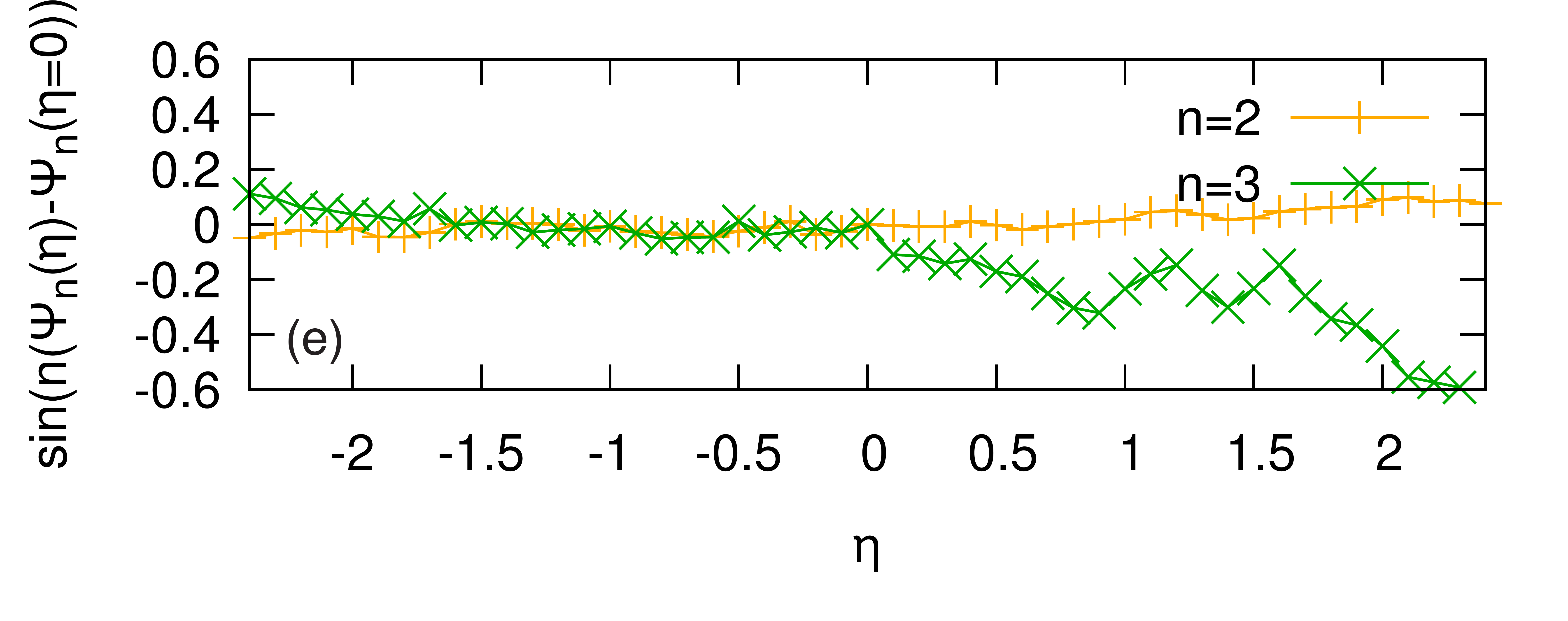} \includegraphics[width=\linewidth]{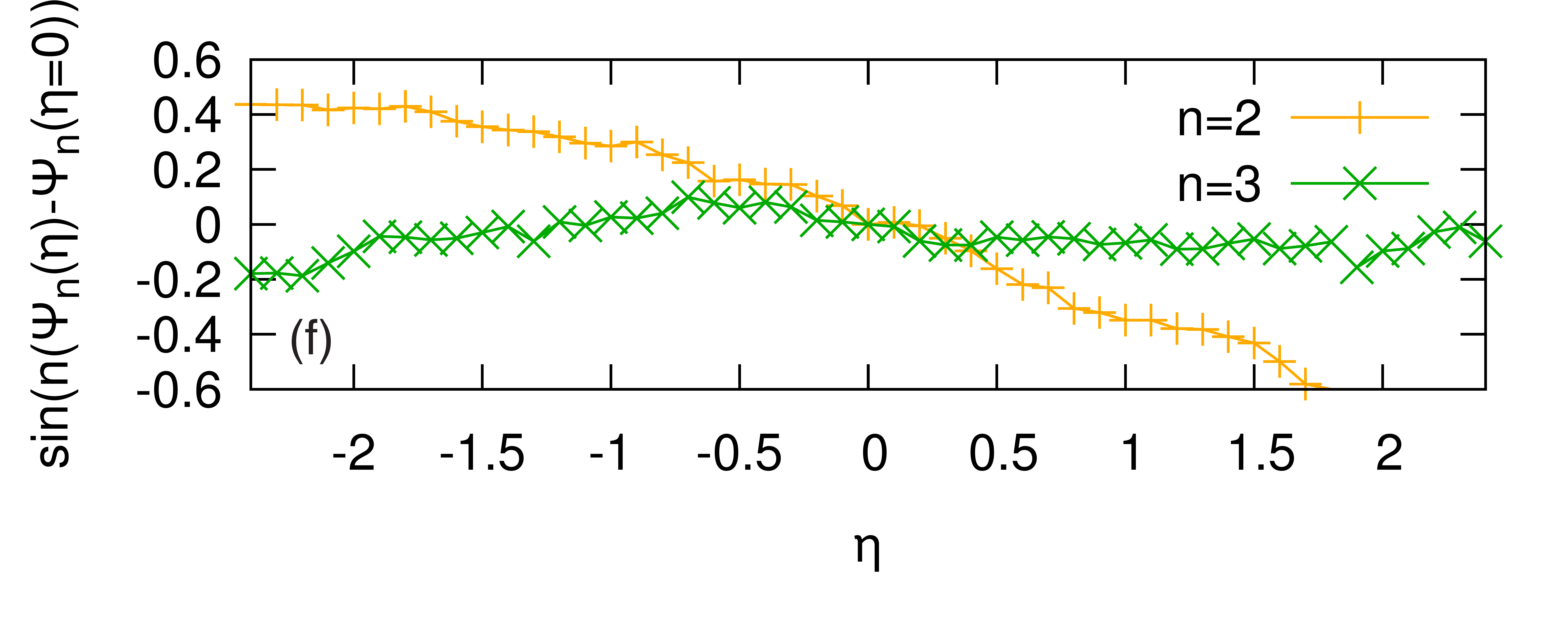}
      \end{minipage}
\end{minipage}
\caption{Change of the eccentricity relative to its value at $\eta=0$ (a-c) and the corresponding variation in angle quantified by $\sin[n(\phi_n(\eta)-\phi_n(0))]$ (d-f) in three single configurations. Simulations used $\alpha_s=0.3$ and $m=0.4\,{\rm GeV}$. \label{fig:singles}}
\end{figure*}

\subsection{3D event geometry}\label{subsec:geometry}
We characterize the transverse geometry by the spatial analogue to the flow $Q$-vector
\begin{equation}\label{eq:evec}
  \boldsymbol{\epsilon}_n(\eta) = \epsilon_n(\eta) e^{in\phi_n(\eta)} = \frac{\int d\mathbf{r}^2 \varepsilon(r,\phi,\eta) r^n e^{i n \phi} }{\int d\mathbf{r}^2 \varepsilon(r,\phi,\eta) r^n }\,,
\end{equation}
defining both the rapidity dependent magnitude $\epsilon_n(\eta)$ and orientation $\phi_n(\eta)$ of the spatial $n$-th order eccentricity. Instead of determining the energy density $\varepsilon$ from the eigensystem of $T^{\mu\nu}$, we neglect the effect of flow velocities in the following and approximate $\varepsilon$ by $T^{\tau\tau}$.

We first demonstrate how the geometry can vary with rapidity in a single event, by showing the change of the eccentricity relative to its value at $\eta=0$ and the corresponding variation in angle quantified by $\sin[n(\phi_n(\eta)-\phi_n(0))]$ for a selection of three typical events in Fig.\,\ref{fig:singles}. Some events show $\epsilon_2$ and $\epsilon_3$ decreasing (or increasing) together, others have them vary in opposite directions.
We do not observe a strong correlation between the variation in angle relative to the change in magnitude of $\epsilon_n$.

Because of its direct relation to the experimental observable used to characterize the decorrelation of anisotropic flows \cite{Khachatryan:2015oea}, we first study 
\begin{align}\label{eq:rn}
  &r_n(\eta_a, \eta_b) = \frac{ \langle {\rm Re}[\boldsymbol{\epsilon}_n(-\eta_a) \cdot  \boldsymbol{\epsilon}^*_n(\eta_b)] \rangle}{ \langle {\rm Re}[\boldsymbol{\epsilon}_n(\eta_a) \cdot  \boldsymbol{\epsilon}^*_n(\eta_b)] \rangle}\notag\\
&= \frac{ \langle \epsilon_n(-\eta_a) \epsilon_n(\eta_b) \cos[n(\phi_n(-\eta_a)-\phi_n(\eta_b))] \rangle}{ \langle \epsilon_n(\eta_a) \epsilon_n(\eta_b) \cos[n(\phi_n(\eta_a)-\phi_n(\eta_b))] \rangle}\,,
\end{align}
where $\boldsymbol{\epsilon}_n$ replaces the flow Q-vector used in the experimental analysis. The brackets $\langle \cdot \rangle$ denote the average over configurations. It was found that the magnitude of $r_n$ in coordinate space  (\ref{eq:rn}) is very close to the decorrelation of final charged hadrons in pseudo-rapidity using a hydrodynamic model for $n=2$ \cite{Pang:2015zrq}. For $n=3$ this correspondence is also good for the central collisions that we study here.

\begin{figure*}[tb]
 \begin{minipage}{\linewidth}
      \centering
      \begin{minipage}{0.46\linewidth}
              \includegraphics[width=\textwidth]{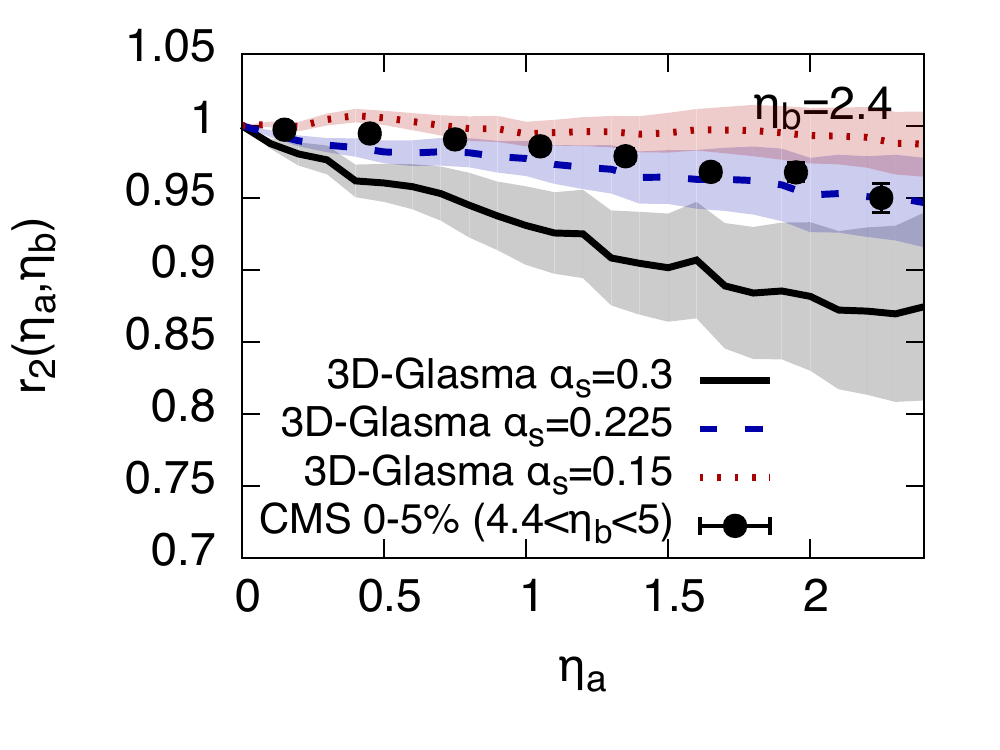}
              \caption{Decorrelation of the initial spatial eccentricity $r_2(\eta_a,\eta_b)$ for $\eta_b=2.4$ in central events ($b=0\,{\rm fm}$) using three different values of $\alpha_s$. We compare to experimental data from the CMS Collaboration \cite{Khachatryan:2015oea}.  
                \label{fig:r2}}
      \end{minipage}
      \hspace{0.05\linewidth}
      \begin{minipage}{0.46\linewidth}
        \includegraphics[width=\textwidth]{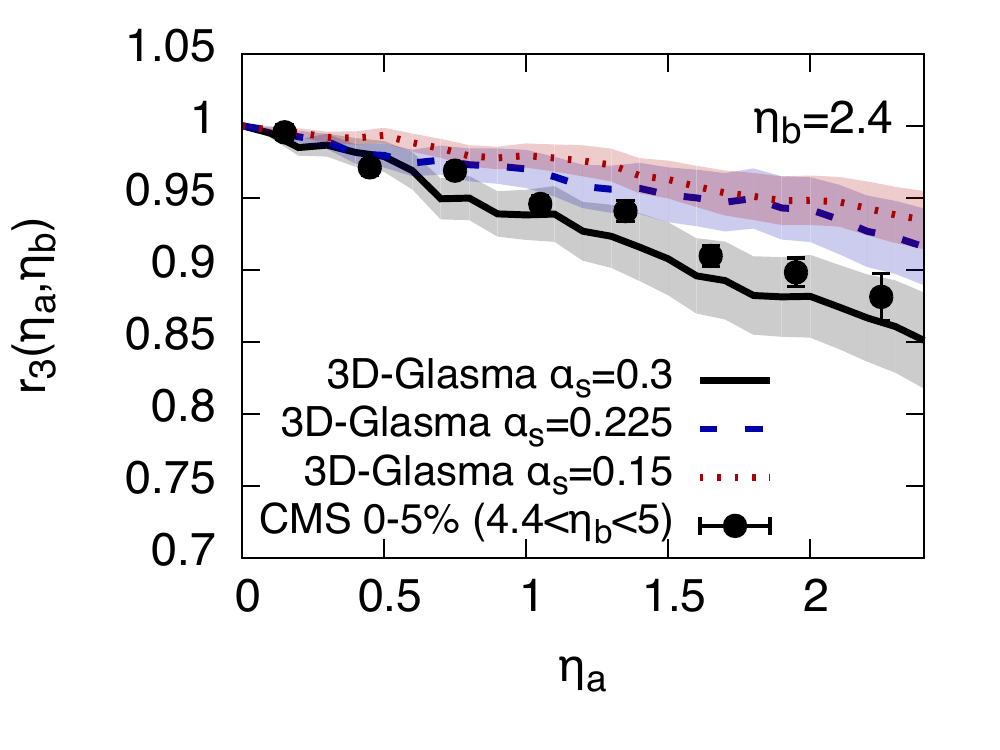}
        \caption{Decorrelation of the initial spatial eccentricity $r_3(\eta_a,\eta_b)$ for $\eta_b=2.4$ in central events ($b=0\,{\rm fm}$) using three different values of $\alpha_s$. We compare to experimental data from the CMS Collaboration \cite{Khachatryan:2015oea}.
          \label{fig:r3}}
      \end{minipage}
\end{minipage}
\end{figure*}

\begin{figure*}[tb]
 \begin{minipage}{\linewidth}
      \centering
      \begin{minipage}{0.45\linewidth}
              \includegraphics[width=\textwidth]{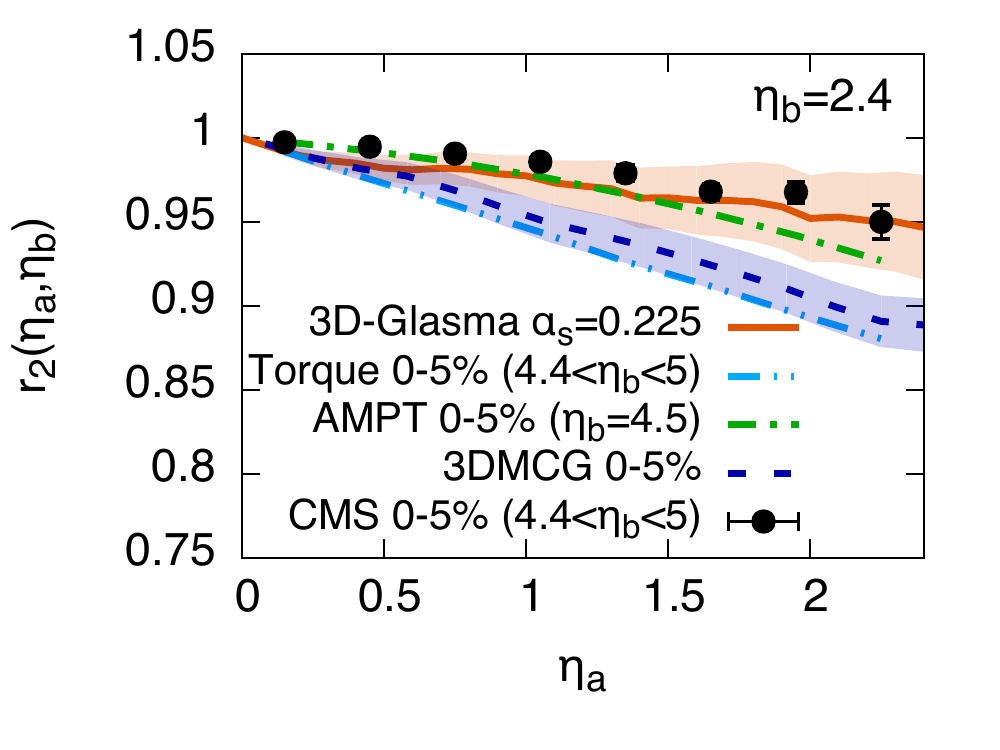}
              \caption{Decorrelation of the initial spatial eccentricity $r_2(\eta_a,\eta_b)$ for $\eta_b=2.4$ in central events ($b=0\,{\rm fm}$) with $\alpha_s=0.225$, comparing to results from the torque model \cite{Bozek:2015bna}, AMPT \cite{Pang:2015zrq} and the 3DMCG model \cite{Monnai:2015sca}. We compare to experimental data from the CMS Collaboration \cite{Khachatryan:2015oea}.  
                \label{fig:r2comp}}
      \end{minipage}
      \hspace{0.05\linewidth}
      \begin{minipage}{0.45\linewidth}
        \includegraphics[width=\textwidth]{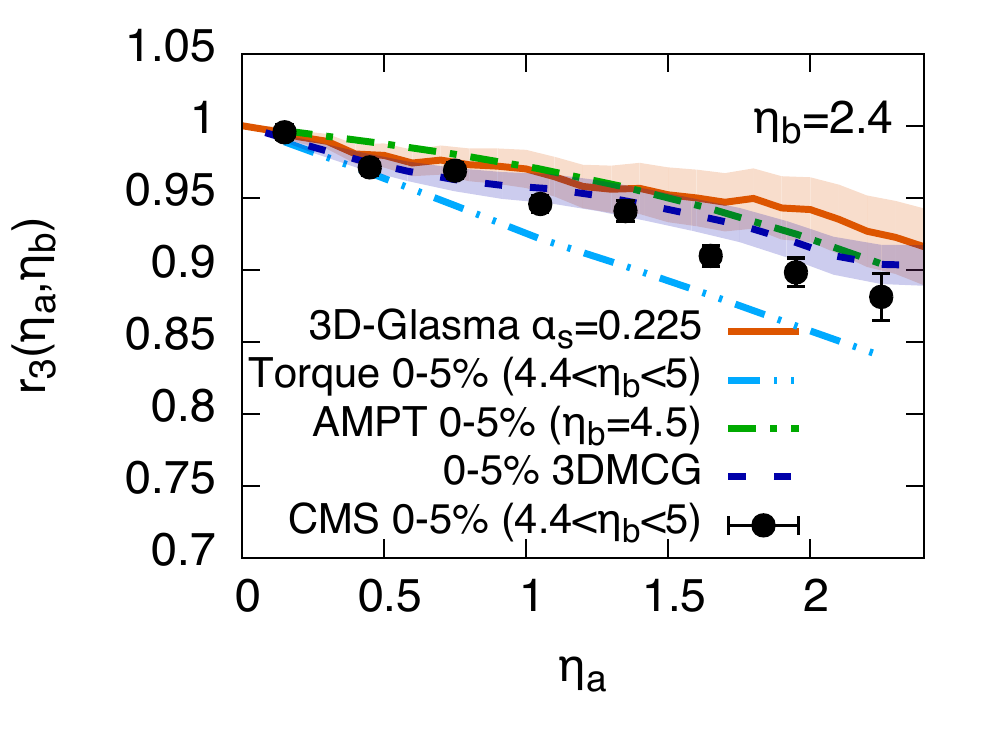}
        \caption{Decorrelation of the initial spatial eccentricity $r_3(\eta_a,\eta_b)$ for $\eta_b=2.4$ in central events ($b=0\,{\rm fm}$) with $\alpha_s=0.225$, comparing to results from the torque model \cite{Bozek:2015bna}, AMPT \cite{Pang:2015zrq} and the 3DMCG model \cite{Monnai:2015sca}. We compare to experimental data from the CMS Collaboration \cite{Khachatryan:2015oea}.
          \label{fig:r3comp}}
      \end{minipage}
\end{minipage}
\end{figure*}

The results for central collisions ($b=0\,{\rm fm}$) with fixed $\eta_b=2.4$, $m=0.4\,{\rm GeV}$ and three different values of $\alpha_s$ are shown in Figs.\,\ref{fig:r2} and \ref{fig:r3}. Since larger values of the coupling constant lead to a faster rapidity evolution of the Wilson lines, they also result in a faster decorrelation of the event eccentricities across different rapidities. However, our results for $r_n$ also show some dependence on the effective mass scale $m$, regulating the growth of the nucleus in impact parameter space. While we find no significant change for $n=2$ when using a smaller value of $m=0.2\,{\rm GeV}$ (and $\alpha_s=0.3$), we find an approximately two times faster drop of $r_3$ with rapidity $\eta_a$ for the smaller value of $m=0.2\,{\rm GeV}$ (and $\alpha_s=0.3$). This can be explained by the increased sensitivity of $r_3$ to the edges of the nuclei ($r^3$ weight in Eq.\,(\ref{eq:evec})), which exhibit a faster rapidity evolution for smaller values of $m$.

We further compare our results to experimental data from the CMS Collaboration \cite{Khachatryan:2015oea}. Given the correspondence of initial and final state $r_n$ values demonstrated in \cite{Pang:2015zrq}, a direct comparison of our results to the experimental data is possible. The experimental data shows a decorrelation for both $n=2$ and $n=3$ that is closest to our result with $\alpha_s=0.225$ and $m=0.4\,{\rm GeV}$. As discussed above, the smaller values of $\alpha_s$ also lead to evolution speeds more compatible with DIS data, and thus should be considered the more realistic choice also for this observable. 

In Figs. \ref{fig:r2comp} and \ref{fig:r3comp}, we compare the 3D-Glasma result for $r_2$ and $r_3$ using $\alpha_s=0.225$ to results from the torque model \cite{Bozek:2015bna}, the 3D Monte Carlo Glauber model (3DMCG) \cite{Monnai:2015sca}, and the AMPT (A Multi-Phase Transport Model) initial conditions studied in \cite{Pang:2015zrq}. 
\begin{figure*}[tb]
 \begin{minipage}{\linewidth}
      \centering
      \begin{minipage}{0.45\linewidth}
        \includegraphics[width=\textwidth]{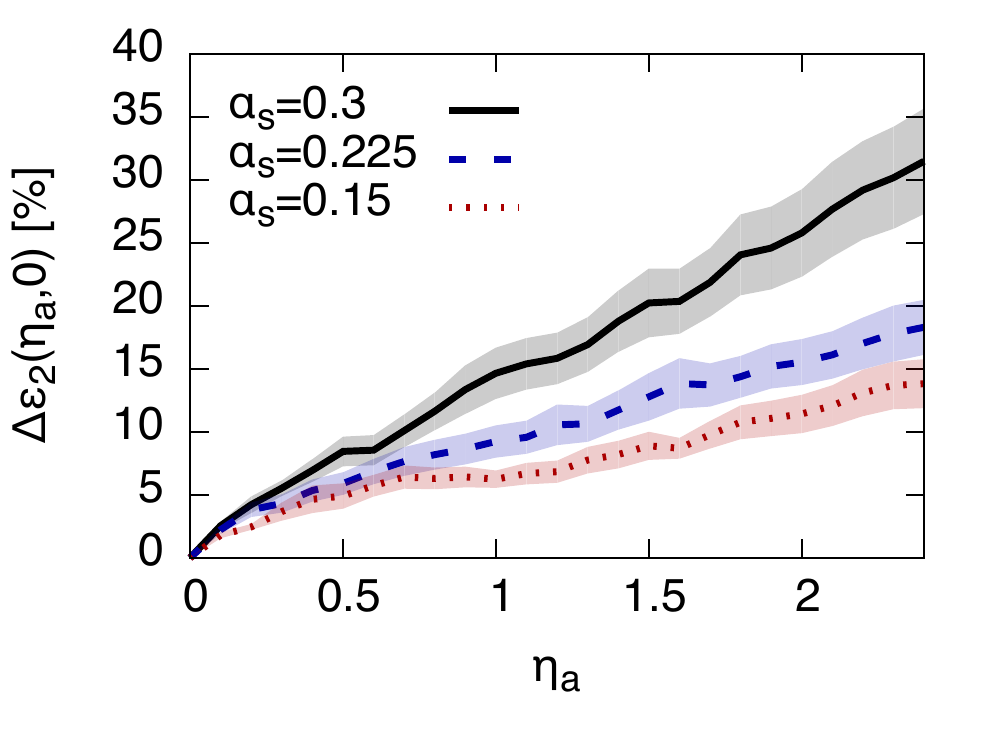}
        \includegraphics[width=\textwidth]{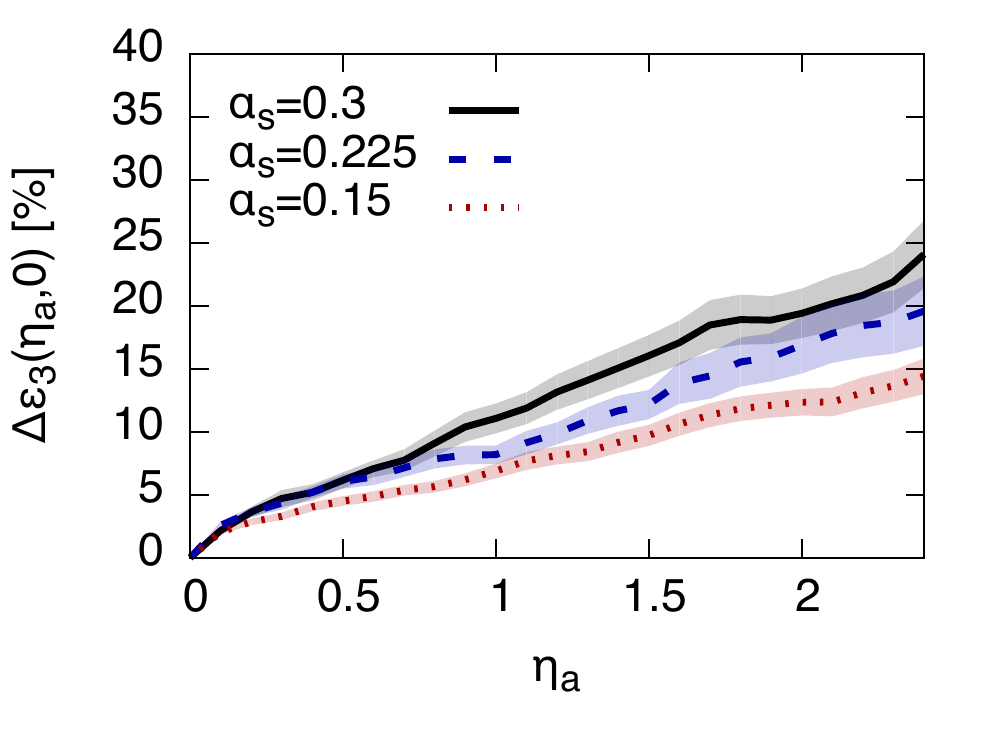}
        \caption{The deviation in percent of $\epsilon_n(\eta_a)$ from $\epsilon_n(0)$ given by $\Delta\epsilon_n(\eta_a,0)$ for $n=2$ (upper panel) and $n=3$ (lower panel) with $m=0.4\,{\rm GeV}$.
          \label{fig:ecc}}
      \end{minipage}
      \hspace{0.05\linewidth}
      \begin{minipage}{0.45\linewidth}
        \includegraphics[width=\textwidth]{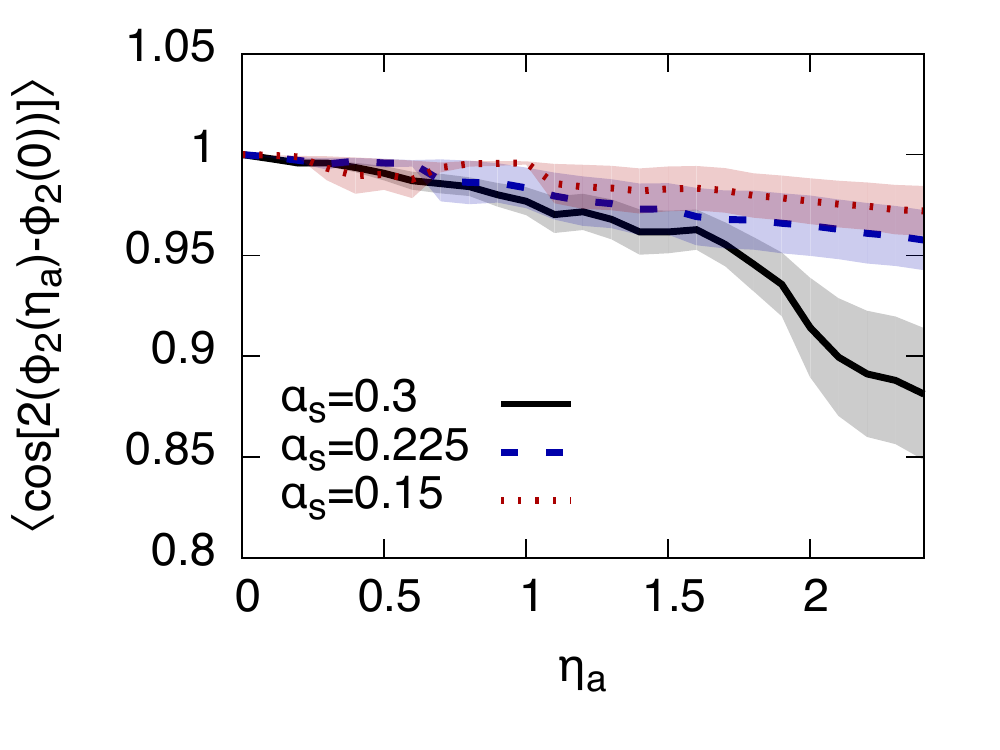}
        \includegraphics[width=\textwidth]{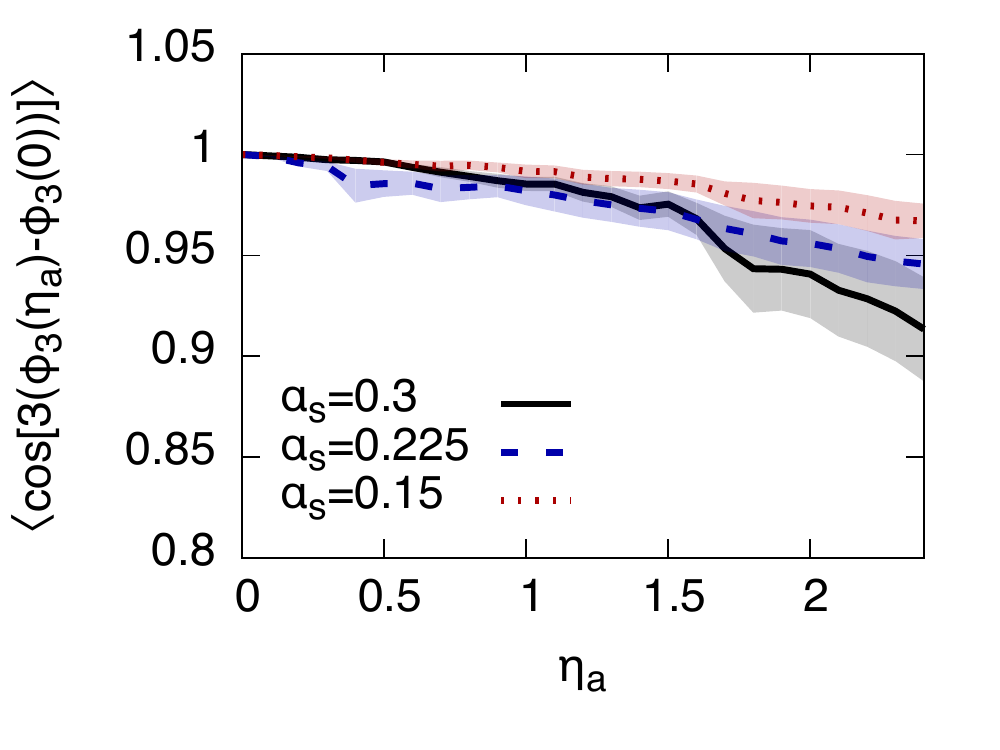}
        \caption{Change in angle $\phi_n(\eta_a)$ relative to $\phi_n(0)$ quantified by $\left\langle \cos [n(\phi_n(\eta_a)-\phi_n(0))] \right\rangle$ for $n=2$ (upper panel) and $n=3$ (lower panel) with $m=0.4\,{\rm GeV}$.
          \label{fig:cos}}
      \end{minipage}
\end{minipage}
\end{figure*}

While in our framework the rapidity dependence is due to quantum fluctuations of color charges in both nuclei described by QCD evolution equations, the other models employ more phenomenological approaches. The 3DMCG model's rapidity dependence results from the fact that the ends of flux tubes between participant quarks follow a given random distribution in rapidity. The torque model generates the effect from asymmetric source profiles in pseudorapidity, and fluctuations of the length of strings. The AMPT model uses HIJING \cite{Wang:1991hta}, which includes fluctuations of string lengths as well as an asymmetric distribution between forward and backward going participants, which depends on the transverse position. The three dimensional spatial structure should thus differ quite significantly between these three models and our framework, yet, the shape of $r_n$ does not seem to be able to distinguish between the models. However, the magnitude of the decorrelation can deviate quite significantly between the models, with our results with $\alpha_s=0.225$ for both $r_2$ and $r_3$ being closest to the AMPT model, and results for $\alpha_s=0.15$ producing weaker decorrelation than any other model.

 It is desirable to find an experimental observable that is more sensitive to how the geometry evolves and fluctuates with rapidity. In the following we analyze in more detail the rapidity dependence of the geometry in the 3D-Glasma framework.

To quantify the typical change of the magnitude of the eccentricity $\epsilon_n(\eta)$ with rapidity, we compute the quantity
\begin{equation}
  \Delta\epsilon_n(\eta_a,\eta_b)=\left\langle\frac{|\epsilon_n(\eta_a)-\epsilon_n(\eta_b)|}{\epsilon_n(\eta_b)}\right\rangle\,.
\end{equation}
The result for fixed $\eta_b=0$ is shown in Fig.\,\ref{fig:ecc}. The deviation grows approximately linearly with $\Delta \eta = |\eta_a-\eta_b|$. We further find approximate scaling with $\alpha_s \Delta\eta$. The slope of the deviation $\Delta\epsilon_n(\eta_a,\eta_b)$ is thus given by $\pm 0.4\,\alpha_s$ per unit of rapidity.

We quantify how the orientation of the eccentricities changes with rapidity using $\left\langle \cos [n(\phi_n(\eta_a)-\phi_n(\eta_b))] \right\rangle$, because this is the relevant quantity entering $r_n$ (\ref{eq:rn}). The result for $\alpha_s=0.15$, 0.225, and 0.3 for $\eta_b=0$ is shown in Fig.\,\ref{fig:cos}.  
The weak dependence of $r_2$ on rapidity (Fig.\,\ref{fig:r2}) can be mainly attributed to the weak change of the orientation shown in Fig.\,\ref{fig:cos}.

\begin{figure}[tb]
\includegraphics[width=0.45\textwidth]{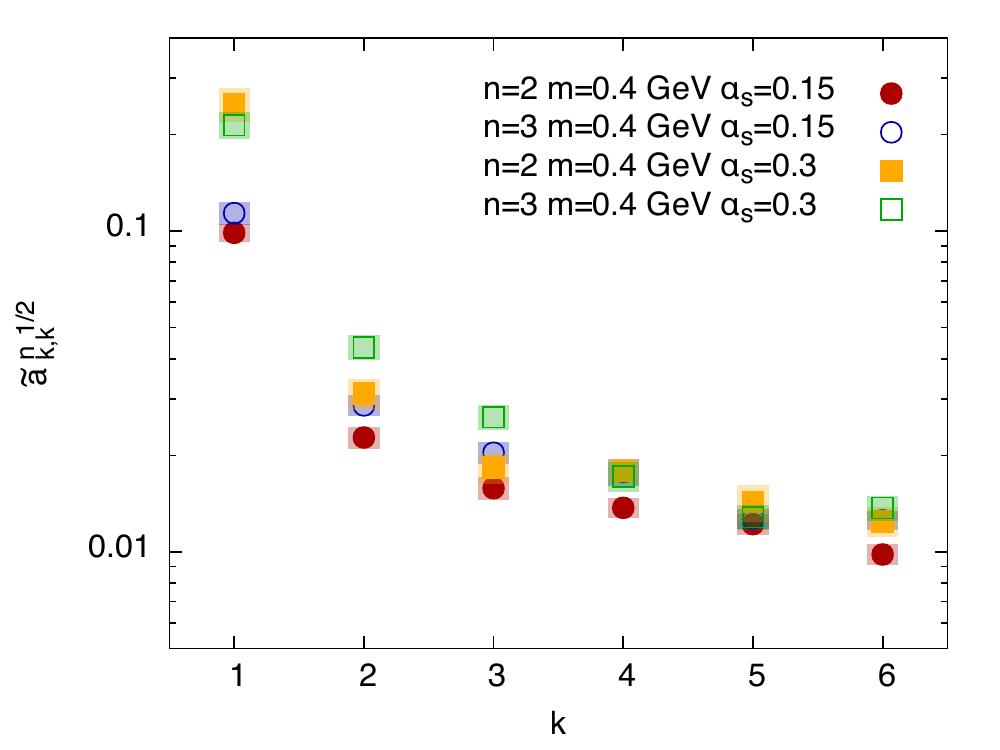}
\caption{Legendre coefficients of the expansion of the eccentricity-vector correlator, characterizing fluctuations of the transverse geometry in rapidity.
\label{fig:anm-ecc}}
\end{figure}

Finally we perform an analysis analogous to that of the multiplicity correlations in Section \ref{subsec:mult} with the eccentricities $\boldsymbol{\epsilon}_2$ and $\boldsymbol{\epsilon}_3$. We define
\begin{equation}\label{eq:Ctilde}
  \tilde{C}^n(Y_1,Y_2) = \frac{\langle {\rm Re}[\boldsymbol{\epsilon}_n(Y_1)\cdot\boldsymbol{\epsilon}^*_n(Y_2)] \rangle}{\langle \epsilon_n(Y_1) \rangle \langle \epsilon_n(Y_2) \rangle}\,,
\end{equation}
compute $\tilde{C}^n_N$ analogous to Eq.\,(\ref{eq:cn}), and perform the same expansion into Legendre polynomials as with $C$ in Eq.\,(\ref{eq:legendre}). Due to limited statistics we constrain this analysis to $\tilde{a}^n_{k,k}=\langle \tilde{a}^n_k \tilde{a}^n_k \rangle$, where $\tilde{a}^n_k$ represents the $k$-th Legendre coefficient of the $n$-th order eccentricity correlator $\tilde{C}^n_N$. We present the results for $m=0.4\,{\rm GeV}$ in Fig.\,\ref{fig:anm-ecc}. No significant $m$ dependence was observed. One can clearly see that also the fluctuations of the transverse geometry in rapidity depend strongly on the evolution speed characterized by $\alpha_s$. It will be very interesting to compare these results to other initial state models and experimental measurements, where $\boldsymbol{\epsilon}_n$ in Eq.\,(\ref{eq:Ctilde}) should be replaced by $Q$-vectors.

\section{Conclusions and Outlook}\label{sec:conc}
We have presented the first step towards a fully three dimensional initial state for heavy ion collisions from quasi first principles calculations within the effective theory of the color glass condensate. Our calculations are accurate for single inclusive quantities to leading logarithmic order in $1/x$. They should further capture the main effects for multi-particle observables that involve different rapidities, in particular for rapidity separations $\lesssim \alpha_s^{-1}$.
We computed the rapidity distributions of produced gluons and their fluctuations, as well as the spatial geometry of the energy momentum tensor and its variation in rapidity. For observables that allow for an approximate comparison to experimental heavy ion data, good agreement is found when using evolution speeds in $x$ that are similar to those extracted from DIS measurements. Interestingly, for this evolution speed our model shows a weaker decorrelation than other models in central collisions, leading to a good description of the usually under-estimated decorrelation measure $r_2$.

This work provides the basis for important phenomenological applications and further theoretical developments. On the phenomenology side, the initial energy momentum tensors computed in this work can in principle be used to initialize viscous hydrodynamic simulations. However, additional modeling will be required to extend the distributions to large rapidities where large $x$ effects, that are not captured in our framework, play a role.
Concerning theoretical improvements, we discussed that at NLO (beyond leading logarithmic accuracy), quantum fluctuations beyond the logarithmically enhanced contribution need to be taken into account. When doing so, full 3-D Yang Mills simulations can be performed, which will be an important next step towards a fully 3-D initial state from first principles.

\vspace{0.2cm}
\section*{Acknowledgments}
We thank Piotr Bozek, Wojciech Broniowski, Tuomas Lappi, Long-Gang Pang, Prithwish Tribedy, and Raju Venugopalan for valuable discussions.
BPS and SS are supported under DOE Contract No. DE-SC0012704. This research used resources of the National Energy Research Scientific Computing Center, which is supported by the Office of Science of the U.S. Department of Energy under Contract No. DE-AC02-05CH11231. BPS acknowledges a DOE Office of Science Early Career Award. SS gratefully acknowledges a Goldhaber Distinguished Fellowship from Brookhaven Science Associates.

\bibliography{spires}

\end{document}